\documentclass[iop]{emulateapj}
\usepackage{apjfonts}

\usepackage{graphicx}

\shorttitle{XRBs across Cosmic Time}
\shortauthors{T. Fragos et al.}

\begin{document}

\title{X-ray Binary Evolution Across Cosmic Time}

\author{T.\ Fragos$^{1}$, B.\ Lehmer$^{2,3}$, M.\ Tremmel$^{4}$, P.\ Tzanavaris$^{2,3}$, A.\ Basu-Zych$^{3}$,  K.\ Belczynski$^{5,6}$,  A.\ Hornschemeier$^{3}$,  L.\ Jenkins$^{3}$, V.\ Kalogera$^{7}$, A.\ Ptak$^{3}$, A.\ Zezas$^{8,9,1}$,} 

\altaffiltext{1}{Harvard-Smithsonian Center for Astrophysics, 60 Garden Street, Cambridge, MA 02138 USA}
\altaffiltext{2}{The Johns Hopkins University, Homewood Campus, Baltimore, MD 21218, USA}
\altaffiltext{3}{NASA Goddard Space Flight Centre, Code 662, Greenbelt, MD 20771, USA}
\altaffiltext{4}{Department of Astronomy, University of Washington, Box 351580, U.W., Seattle, WA 98195-1580, USA}
\altaffiltext{5}{Astronomical Observatory, University of Warsaw, Al. Ujazdowskie 4, 00-478 Warsaw, Poland}
\altaffiltext{6}{Center for Gravitational Wave Astronomy, University of Texas at Brownsville, Brownsville, TX 78520, USA}
\altaffiltext{7}{Department of Physics and Astronomy, Northwestern University, 2145 Sheridan Road, Evanston, IL 60208, USA}
\altaffiltext{8}{Department of Physics, University of Crete, P.O. Box 2208, 71003 Heraklion, Crete, Greece}
\altaffiltext{9}{IESL, Foundation for Research and Technology, 71110 Heraklion, Crete, Greece}
\email{tfragos@cfa.harvard.edu}
%, mjt29@astro.washington.edu, blehmer@pha.jhu.edu, panayiotis.tzanavaris-1@nasa.gov, kbelczyn@astrouw.edu.pl, vicky@northwestern.edu, antara.r.basu-zych@nasa.gov,  leigh.jenkins-1@nasa.gov, Ann.Hornschemeier@nasa.gov, andrew.f.ptak@nasa.gov, azezas@physics.uoc.gr}

\begin{abstract}
	
High redshift galaxies permit the study of the formation and evolution of X-ray binary populations on cosmological timescales, probing a wide range of metallicities and star-formation rates. In this paper, we present results from a large scale population synthesis study that models the X-ray binary populations from the first galaxies of the universe until today. We use as input to our modeling the Millennium II Cosmological Simulation and the updated semi-analytic galaxy catalog by \citet{Guo2011} to self-consistently account for the star formation history and metallicity evolution of the universe. Our modeling, which is constrained by the observed X-ray properties of local galaxies, gives predictions about the global scaling of emission from X-ray binary populations with properties such as star-formation rate and stellar mass, and the evolution of these relations with redshift. Our simulations show that the X-ray luminosity density (X-ray luminosity per unit volume) from X-ray binaries in our Universe today is dominated by low-mass X-ray binaries, and it is only at $z \gtrsim 2.5$ that high-mass X-ray binaries become dominant. We also find that there is a delay of $\sim 1.1\,\rm Gyr$ between the peak of X-ray emissivity from low-mass Xray binaries (at $z\sim 2.1$) and the peak of star-formation rate density (at $z\sim 3.1$). The peak of the X-ray luminosity from high-mass X-ray binaries (at $z\sim 3.9$), happens $\sim 0.8\,\rm Gyr$ before the peak of the star-formation rate density, which is due to the metallicity evolution of the Universe.

\end{abstract}


\maketitle

\section{INTRODUCTION}

X-ray observations of galaxies are a key component to unraveling the physics of compact object formation. Stellar remnants and the distribution of the hot interstellar medium are best studied via their X-ray emission. With {\em Chandra} several global galaxy X-ray correlations have been established and appear to hold to at least $z=1$ \citep[e.g.][]{Bauer2002,Lehmer2008,Vattakunnel2012,Symeonidis2011,Lehmer2012}, implying a large-scale regularity to the production of X-ray binaries (XRBs) and hot gas, the primary sources of high energy emission in ``normal'' (non-AGN) galaxies. These include a very strong correlation between total X-ray emission from high-mass XRBs (HMXBs) and the galaxy-wide star-formation rate (SFR) \citep[e.g. ][]{Ranalli2003,Gilfanov2004a,Hornschemeier2005,Lehmer2010, Mineo2012a}, as well as a scaling between total X-ray emission from low-mass XRBs (LMXBs) and stellar mass ($M_*$) \citep{Gilfanov2004b,Lehmer2010,Boroson2011,ZGA2012}. These correlations present exciting challenges for models of accreting binary evolution \citep[e.g.][]{Fragos2008, Fragos2009, Belczynski2008} that predict significant variability in these relations based on, e.g., star formation history and metallicity. 

Recent ultradeep Chandra and multiwavelength surveys \citep[e.g. ][]{Xue2011} have permitted robust observational tests of these correlations in distant galaxies \citep[see also the review of][]{BH2005}. In a broad sense, curiously, the X-ray emission per unit SFR seems to evolve little with redshift, despite the evolution in gas metallicity, for example, over the same epochs \citep[e.g.][]{Lehmer2008}. \citet{Ptak2007} and \citet{TG2008} used rich multi-wavelength datasets to construct X-ray luminosity functions (XLFs) of normal galaxies in different redshift bins, and found that the XLF shows a statistically significant evolution with redshift and that it is the late type galaxies that are driving this evolution.

Despite the significant investment of observing time in X-ray studies of distant normal galaxies, and the recent theoretical advances in modeling the X-ray source populations in nearby galaxies, our understanding of the cosmological evolution of populations of compact objects is still in its infancy. The first steps in this direction have been made by means of semi-analytic, empirical models \citep{WG1998,GW2001}. Considering a time-dependent SFR, these models predicted that the time required for binaries to reach the X-ray phase (due to the donor's nuclear evolution and angular momentum losses) leads to a significant time delay between a star-formation episode and the production of X-ray emission from X-ray binaries from this populations. 

State-of-the-art binary population models \citep[e.g.][]{Belczynski2008} provide us with a more physical picture of XRB populations as a function of galaxy properties. This type of detailed modeling has been applied to both starburst and elliptical nearby galaxies, where large populations of individual XRBs are resolved, and detailed ages, star-formation histories, and metallicities are measured. \citet{Belczynski2004} constructed the first synthetic XRB populations for direct comparison with the Chandra observed XLFs of NGC\,1569, a star forming dwarf irregular galaxy. \citet{Linden2009, Linden2010} developed models for the HMXB and Be XRBs of SMC, studying the XLF and the spatial distribution of this population, and investigating the effect of electron-capture supernovae of massive ONeMg stellar cores. \citet{Fragos2008, Fragos2009} performed extensive population synthesis (PS) simulations,  modeling the two old elliptical galaxies NGC\,3379 and NGC\,4278, constraining the relative contribution to the observed XLF from sub-populations of LMXBs with different donor and accretor types, and the effects of the transient behavior of LMXBs. More recently, \citet{ZL2011} studied the evolution of XRB populations in late type galaxies, using a modified version of the BSE population synthesis code \citep{Hurley2002}. Considering different star-formation history scenarios, they were able to derive the X-ray emissivity as well as XLFs of the XRB populations. However, their model predictions are not in good agreement with observations for most of the Universe lifetime: from $4.5$ to $13.7\,\rm Gyr$ ($z=0-1.4$).   

The next step in this long-term effort is to apply these models in a cosmological context, in order to understand the nature of the accreting populations in high-redshift normal galaxies. In this paper, we study the global scaling relation of emission from XRB populations with properties such as SFR and stellar mass, and the evolution of these relations with redshift. More specifically, we developed a large grid of XRB PS models for which the information about the star-formation history and metallicity evolution were derived from cosmological simulations. Our models, which we have already constrained with observations of XRB populations of the local universe, give predictions for the evolution of the aforementioned global scaling relations back to the formation of the very first galaxies at redshift $z \gtrsim 15$.

The plan of the paper is as follows. In Section~2 we describe our simulation tools, the {\tt StarTrack} PS code and Millennium II simulation, and the methodology we follow in developing models for the evolution of XRBs on cosmological timescales. Section~3 discusses the necessary bolometric corrections we employ in order to directly compare our model results with observations. In Section~4, we present the observational data we use to constrain our models and the statistical analysis we follow. We present the predictions of our maximum likelihood models for the evolution of XRB population in the high redshift universe in Section~5. Finally, in Section~6 we summarize the main findings and conclusions of our work. 

%%%%%%%%%%%%%%%%%%%%%%%%%%%%%%%%%%%%%%%%%%%%%%%%%%%%%%%%%%%%%%%%%%%%%%%%%%%%%
%%%%%%%%%%%%%%%%%%%%%%%%%%%%%%%%%%%%%%%%%%%%%%%%%%%%%%%%%%%%%%%%%%%%%%%%%%%%%
\section{Population Synthesis Modeling}

%%%%%%%%%%%%%%%%%%%%%%%%%%%%%%%%%%%%%%%%%%%%%%%%%%%%%%%%%%%%%%%%%%%%%%%%%%%%%
\subsection{Population Synthesis Code: {\tt StarTrack}}

The main tool we use to perform the PS simulations is {\tt StarTrack} \citep{Belczynski2002,Belczynski2008}, a state-of-the-art  code with special emphasis on processes leading to the formation and further evolution of compact objects. Both single and binary star populations are considered. The code incorporates calculations of all mass-transfer phases, a full implementation of orbital evolution due to tides, as well as estimates of magnetic braking. {\tt StarTrack} has been extensively tested and calibrated using detailed mass-transfering binary star calculations and observations of binary populations. For a comprehensive summary of {\tt StarTrack} and its model parameters see  \citet{Belczynski2008}

Recently the {\tt StarTrack} code has undergone three major revisions. The first  one allows for the updated stellar winds and their re-calibrated dependence on  metallicity \citep{Belczynski2010}. This revision is fully incorporated into our results. The second update, based on the fully consistent supernova simulations,  resulted in the revised neutron star and black hole mass spectrum \citep{Belczynski2012,Fryer2012}. Finally, the most recent upgrade involves  a more physical treatment of donor stars in common envelope events via usage of actual value of $\lambda$ parameter for which usually a constant value is assumed \citep{Dominik2012}. The last two code revisions are not employed in our simulations as these were performed well before the revisions became available. 

Several XRB PS studies that used {\tt StarTrack}  \citep{Belczynski2004,Belczynski2004b,Fragos2008, Fragos2009, Fragos2010, Linden2009, Linden2010} have been pivotal in interpreting observations of nearby galaxies and showed that despite the significant number of free parameters ($\sim 10$) and the complexity of the physical processes involved, PS models can provide robust results that can be used to understand the properties of more distant galaxies. 

We should stress here that our PS code only models XRBs formed via the evolution of primordial isolated binaries, i.e. the field XRB population. A dynamically formed population of LMXBs can have a significant contribution to the integrated X-ray luminosity of some globular cluster rich elliptical galaxies, where in certain cases more than half of the bright LMXBs reside in globular clusters \citep[e.g.][]{HB2008}. The detailed modeling of the dynamically formed LMXBs, although possibly important for this subset of elliptical galaxies, is outside the scope of this paper, and in fact it is not feasible given the tools that exist to date for simulations of such scale. The only PS studies to date of dynamically formed LMXBs have been done for a few specific Galactic globular clusters and only in order to identify the various formation channels and their relative importance \citep{Ivanova2006, Ivanova2008, Ivanova2010}.

In the work presented here we consider the global XRB population, i.e. XRBs in a general galaxy population of both early and late galaxies, where the contribution of dynamically formed LMXBs is much smaller.  In the low redshift Universe approximately half of the stellar mass is in early type galaxies \citep{Bell2003}. Furthermore, approximately half of early type galaxies are old, globular-cluster rich ellipticals, and finally in these globular-cluster rich ellipticals approximately half of the bright LMXBs are located in globular clusters. This effectively translates to a possible error of $10\%-25\%$ in the determination of the integrated X-ray luminosity coming from the LMXB population when neglecting the dynamically formed population. This contribution of the dynamically formed population is decreasing further as we move to higher redshifts.  However, comparing the X-ray luminosity per unit stellar mass coming from LMXBs in star-forming galaxies \citep{Lehmer2010}, where the dynamically formed LMXB population is negligible, and in elliptical galaxies \citep{Boroson2011}, shows that any correction that one should make due to the contribution of dynamically formed LMXBs is smaller than the observational uncertainties.

%%%%%%%%%%%%%%%%%%%%%%%%%%%%%%%%%%%%%%%%%%%%%%%%%%%%%%%%%%%%%%%%%%%%%%%%%%%%%
\subsection{The Millennium-II Simulation}

The Millennium-II Simulation \citep{Boylan2009} is a very large N-body simulation of dark matter evolution in the concordance $\Lambda$CDM cosmology. The simulation uses $\sim 10^{10}$ particles in a $100 Mpc/h$ box, resulting in a spatial resolution of $1 kpc/h$ and a mass resolution of $6.89 \times 10^6 M_\odot/h$. 67 full snapshots of the simulation were stored, with time intervals between them varying from $\sim 30\,\rm Myr$ to $\sim 300\,\rm Myr$, which were then used to derive merger trees of the dark-matter halos.

Using the stored halo/subhalo merger trees of the Millennium II simulation, \citet{Guo2011} recently released an updated semi-analytic galaxy catalog. \citet{Guo2011} revised the modeling of several physical processes involved in galaxy formation, such as gas stripping and supernova feedback. This newly released catalog yields an excellent fit to the galaxy mass and luminosity function of low-redshift galaxies over five orders of magnitude in stellar mass and nine magnitudes in luminosity. The \citet{Guo2011} catalog provides a wealth of information for each galaxy in the cosmological simulation, including galaxy type, SFR, metallicity, hot and cold gas content, central black hole (BH) mass, and rest-frame absolute magnitudes in the SDSS filters,  as a function of time.

\begin{figure}
\centering
\includegraphics{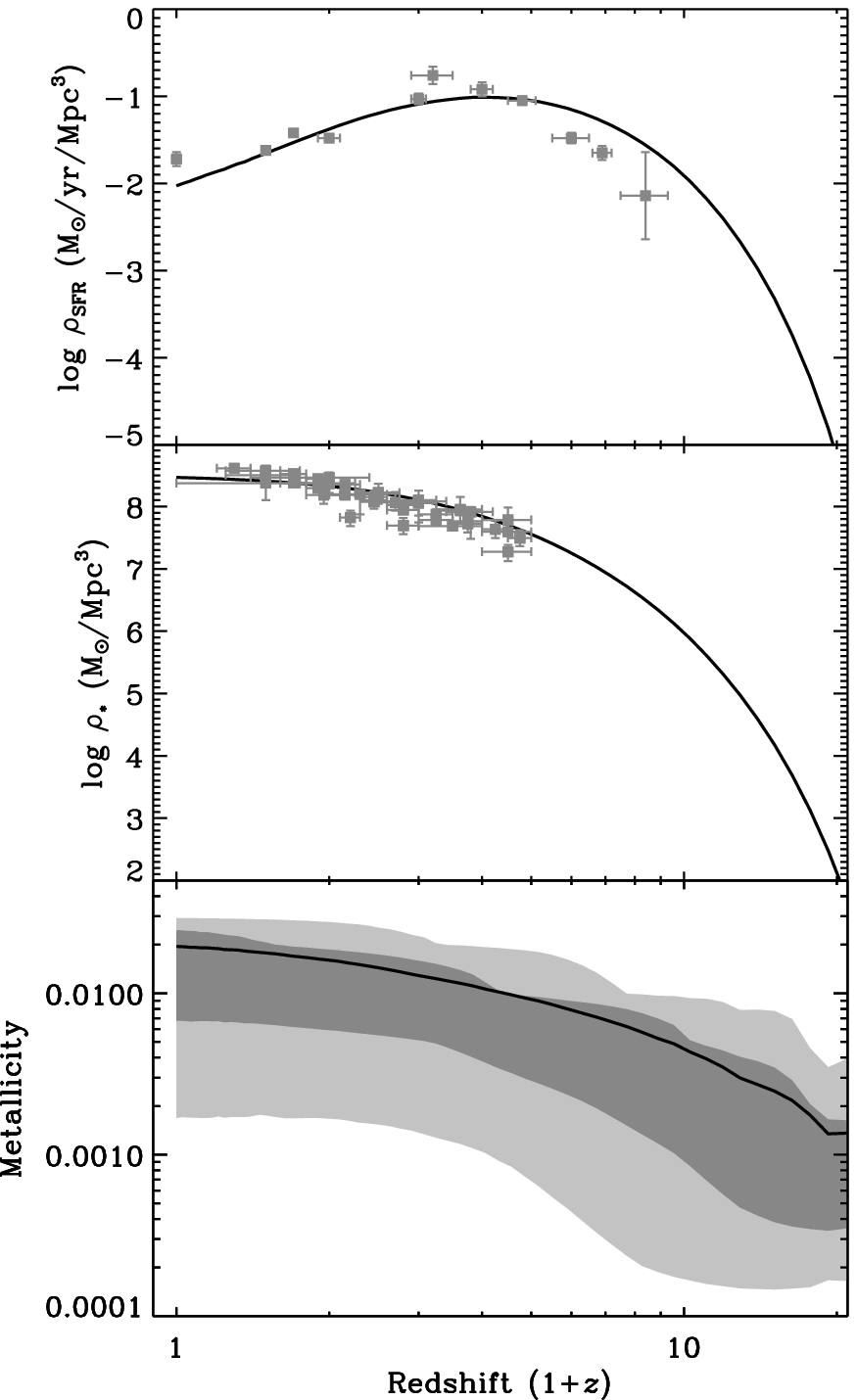}
\caption{\label{MRII} The top panel shows the SFR density evolution with redshift, as predicted by the \citet{Guo2011} modeling (solid line). Observational determinations of the SFR density at different redshifts (grey squares) are shown for comparison \citep{Schiminovich2005, Bouwens2007,RS2009}. The stellar mass density as a function of redshift, as predicted by the the \citet{Guo2011} modeling, is shown in the middle panel (solid line), as well as a compilation of observed data (grey squares) by \citet{Marchesini2009}. The bottom panel shows the mass weighted mean metallicity of the newly formed stellar mass as a function of redshift. The dark grey area shows the metallicity range with which 68.2\% of the new stars form, as a function of redshift, while the light grey area corresponds to the 95.4\% of the newly formed stars.}
\end{figure}

As already mentioned, the Millennium II simulation is a dark matter only simulation, and the baryonic matter is modeled as an add-on, using semi-analytic prescriptions. Despite the high complexity of the galaxy formation modeling by \citet{Guo2011}, which takes into account most physical processes involved, it remains an approximate model. In that sense, the detailed evolution of a specific galaxy formed in the simulation  may not be fully consistent with more advanced galaxy formation simulations \citep{Scannapieco2012}. However, the semi-analytic prescriptions used by \citet{Guo2011} have been calibrated so that the statistical properties (e.g. mass-function, global star-formation history, stellar metallicity distribution) as a function of redshift of the modeled galaxy population are consistent with available observational data. Figure~\ref{MRII} shows the predictions of the Millennium-II simulation and the \citet{Guo2011} modeling for the SFR density (top panel), the stellar mass density (middle panel), and the metallicity of the newly formed stars as a function of redshift. For comparison observed data for the SFR density and stellar mass density are also plotted (upper and middle panel respectively), showing a good agreements between model and observations. 

%%%%%%%%%%%%%%%%%%%%%%%%%%%%%%%%%%%%%%%%%%%%%%%%%%%%%%%%%%%%%%%%%%%%%%%%%%%%%
\subsection{The Millennium-II simulation as Initial Conditions of Populations Synthesis Models}

The first step of our analysis is to derive for each combination of PS model parameters, and each value of metallicity, the evolution of the specific X-ray luminosity (X-ray luminosity per unit stellar mass) of a coeval population of XRBs as a function of its age. Assuming a single instantaneous starburst, we follow the evolution of the XRB population, keeping track of the total (cumulative) X-ray luminosity of the whole XRB population, and the contribution of the different sub-populations (e.g. XRBs with different types of accretors) as a function of the age of the population. We note that in this work we define as a LMXB, a XRB with donor star whose current mass is less than $3\, \rm M_{\odot}$, while a HMXB is any XRB with a donor star more massive than $3\, \rm M_{\odot}$.

\begin{figure}
\centering
\includegraphics{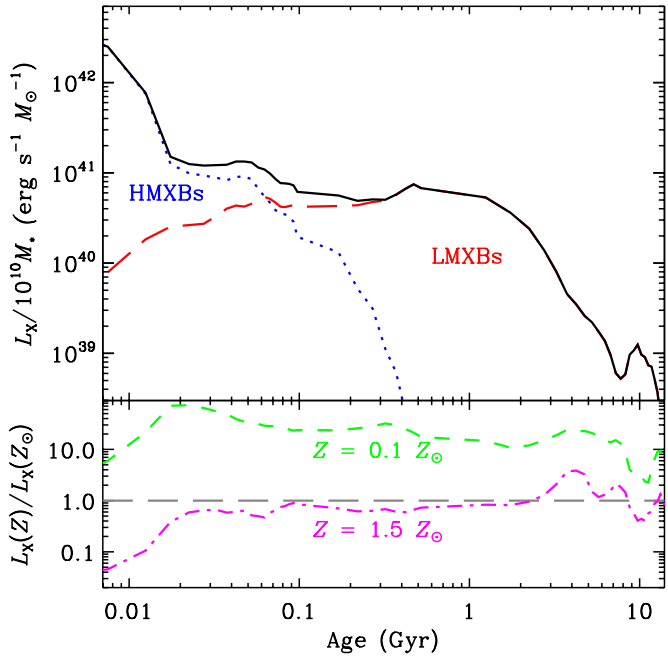}
\caption{\label{single_age} The evolution of an XRB population formed from a single-burst population of stars. \textbf{Top panel:} Specific bolometric X-ray luminosity, i.e. luminosity per unit stellar mass, of a coeval stellar population as a function of the population's age for our ``reference'' model (Model 245; see also sections 2.4 and 4.2) and for solar metallicity, broken down into HMXB (blue dotted) and LMXB (red long-dashed) contributions. After an initial $\sim 5\, \rm Myr$ ramp-up, HMXBs dominate the X-ray power output for $\sim 100-300\, \rm Myr$. Following this, LMXBs take over, peak in formation at $\sim 0.5-1\rm \, Gyr$, and then passively fade with increasing age. The bumps and wiggles present at $\gtrsim 4\,\rm Gyr$ are due to statistical fluctuations as bright sources turn on and off. \textbf{Bottom Panel:} X-ray binary evolution for the cases of metallicities $Z = 0.1Z_{\odot}$ (green short-dashed) and $Z = 1.5Z_{\odot}$ (magenta dot-dashed) compared to solar (i.e., $L_X[Z]/L_X[Z_{\odot}]$).}
\end{figure}

In the top panel of Figure~\ref{single_age}, we show the specific X-ray luminosity of XRBs, X-ray luminosity per unit stellar mass ($L_X/M_{*}$), versus age for a single coeval population of binaries at solar metallicity. These curves were generated using our ``reference'' model (Model 245; see also sections 2.4 and 4.2). In the bottom panel of Figure~\ref{single_age}, we show how these curves vary if instead initial metallicities of $\sim 0.1 Z_\odot$ (green dashed) and $\sim 1.5 Z_\odot$ (magenta dot-dashed) are chosen. As seen in the top panel of Figure~\ref{single_age} over the lifetime of the stars produced in a given star-formation event, there is an initial ramp-up in the formation of HMXBs over the first $\sim 5\rm\, Myr$ as the first compact objects form, followed by a period of HMXB dominance that lasts for $\sim 100-300\rm\, Myr$. At this point, the LMXB population emerges, peaks, and then slowly declines by $\sim 1-2$ orders of magnitude out to $\sim 14\rm\, Gyr$ timescales. In the lower panel of Figure~\ref{single_age} we show the X-ray binary luminosity evolution with respect to solar for low and high metallicity cases. We note that the X-ray luminosity is expected to be significantly higher for low-metallicity regions, due to the fact that the young massive O/B stars, which are the precursors to compact object accretors in X-ray binaries, will lose less stellar mass from line-driven winds over their lifetimes for lower metallicities. This results in more numerous and more massive BH populations, and subsequently more luminous X-ray binary populations (see also section 5.1 for a detailed discussion). Some evidence has already indicated that the X-ray luminosity from XRBs per unit of SFR ($L_X/SFR$) for low-metallicity dwarf galaxies may be enhanced compared to more typical galaxies \citep{KSG2011}.

The Millennium II simulation and the \citet{Guo2011} semi-analytic galaxy catalog provide us information about the evolution of the properties of each galaxy in the simulation. In this first study, we consider only the global characteristics and scaling relations of the XRB evolution across cosmic time. Having this in mind, we consider the whole simulation box of the Millennium II simulation as representative of the Universe, and we keep track of the total new stellar mass that is formed between each snapshot of the simulation. In addition, we keep track of the metallicity of the new stellar mass formed. In practice, we consider 9 metallicity bins, centered at the values reported in Table~\ref{modelparam}, and calculate how much new stellar mass was formed at each snapshot and at each metallicity bin. Finally, we assume that the new stellar mass is formed with a constant SFR between each snapshot. This results in a self-consistent prescription of the star-formation history and metallicity evolution of the Universe.

We now have all the necessary information in hand to construct models for the evolution of XRBs across cosmic time. We convolve the relations we derived in the first step, with the star-formation history for different metallicity ranges. The result is a mixed XRB population, with a broad range of ages and metallicities, representative of the XRB population of the Universe, at each redshift. From this convolution we derive model predictions such as the X-ray luminosity from XRBs per unit volume, stellar mass, or SFR, as a function of redshift.

%%%%%%%%%%%%%%%%%%%%%%%%%%%%%%%%%%%%%%%%%%%%%%%%%%%%%%%%%%%%%%%%%%%%%%%%%%%%%
\subsection{Parameter study}

{\tt StarTrack} PS models have a significant number of free parameters, which can be categorized into two groups. In the first group there are parameters that usually correspond to the distributions of the initial properties of the binary population, such as the initial mass function (IMF), the initial binary mass ratio ($q$), and the distribution of initial orbital separations. For these parameters, we primarily use fixed values derived from the most updated observational surveys. Only in the case that there is an ongoing debate regarding a parameter we actually treat it as a free parameter. In the same group one should add parameters such as the stellar wind prescriptions, and the natal kick distribution of neutron stars (NSs), as these parameters are also derived from observational data. The second group is that of the few ``true'' free parameters that correspond to poorly understood physical processes which we are not able to model in detail. In this second group belong parameters such as those involved in the common envelope (CE) phase modeling. 

We created the largest ever grid of 288 PS models. Each of these 288 models was run at 9 different values of metallicity ($Z=10^{-4}-0.03$), where we followed the evolution of $\sim 45$ million binaries per model. The whole set of simulations required $\sim 2$ million CPU hours of computational time. In this grid, we varied 6 parameters (in addition to the metallicity) known from previous studies \citep{Belczynski2007,Belczynski2010, Fragos2008, Fragos2010, Linden2009} to affect the evolution of XRBs and the formation of compact objects in general. Namely, we varied the CE efficiency ($\alpha_{CE}\times\lambda=0.1-0.5$), the IMF (Kroupa 2001 or Kroupa \& Weidner 2003), the initial mass ratio ($q$) distribution (flat with $q=0-1$, twin with $q=0.9-1.0$, or a mixture of 50\% flat and 50\% twin), the stellar wind strength \citep[$\eta_{wind}=0.25,\,1.0,\, or\, 2.0$; parameter with which we multiply the stellar wind prescription described in][]{Belczynski2010}, and the distribution of natal kicks for BHs formed though direct core collapse (zero kicks, or 10\% of the standard kicks for NS). Finally, we allow for various outcomes of CE phase, in particular taking into account the possible CE inspirals with Hertzsprung gap donors that terminate binary evolution barring the subsequent XRB formation \citep{Belczynski2007}. The grid of 288 models we created includes all the possible combinations of the parameters we varied. The various parameter values used are shown in Table~\ref{model_param}. 

We note that in our calculations we combine $\alpha_{\rm CE}$ and $\lambda$ into one CE parameter, where $\lambda$ is a measure of the central concentration of the donor and the envelope binding energy. In the rest of the text, whenever we mention the CE efficiency $\alpha_{\rm CE}$, we refer in practice to the product $\alpha_{\rm CE} \times \lambda$ \citep[see][for details]{Belczynski2008}.

\begin{deluxetable*}{lccc} 
%\rotate
\centering
\tablecolumns{4}
\tabletypesize{\scriptsize}
\tablewidth{0pt}
\tablecaption{Model Parameters 
\label{modelparam}} 
\tablehead{ \colhead{Parameter} & 
     \colhead{Notation} & 
     \colhead{Value} &
     \colhead{Reference} 
     }
\startdata
Initial Orbital Period distribution & F(P) 				& flat in logP 													& \citet{Abt1983}\\
Initial Eccentricity Distribution 	& F(e)		 		& Thermal $F(e)\sim e$ 											& \citet{Heggie1975}\\
Binary Fraction  					& $f_{\rm bin}$ 	& 50\% 															& \\
Magnetic Braking 					&					& 																& \citet{IT2003} \\
Metallicity           				& $Z$           	& 0.0001, 0.0002, 0.005, 0.001,  								& \\
									&					& 0.002, 0.005, 0.01, 0.02, 0.03								& \\
IMF  (slope) 						&      				&  -2.35 or  -2.7 												& \citet{Kroupa2001,KW2003}\\
Initial Mass Ratio Distribution 	& F(q) 				& Flat, twin, or 50\% flat + 50\% twin 						& \citet{KF2007,PS2006}\\
CE Efficiency 						& $\alpha_{\rm CE}$ &  0.1, 0.2, 0.3, or 0.5 										& \citet{PRH2003}\\
Stellar wind strength 				& $\eta_{\rm wind}$ &  0.25, 1.0, or 2.0 											& \citet{Belczynski2010}\\
CE during HG 						& 					& Yes or No 													& \citet{Belczynski2007}\\
SN kick for ECS/AIC\tablenotemark{a} NS				& 					& 20\% of normal NS kicks	 									& \citet{Linden2009}\\
SN kick for direct collapse BH 		& 					& Yes or No 													& \citet{Fragos2010}

\enddata
\label{model_param}
\tablenotetext{a}{Electron Capture Supernova / Accretion Induced Collapse}
\end{deluxetable*}

%%%%%%%%%%%%%%%%%%%%%%%%%%%%%%%%%%%%%%%%%%%%%%%%%%%%%%%%%%%%%%%%%%%%%%%%%%%%%
\section{Bolometric Corrections to the Chandra X-ray Bands}

{\tt StarTrack}, our PS synthesis code, keeps track of the mass-transfer rate as a function of time for every modeled XRB. From this mass-transfer rate, we derive the bolometric luminosity based on the prescriptions presented by \citet{Fragos2008,Fragos2009}, which also take into account the transient behavior of XRBs. However, in order to compare our models with observational data, we need to estimate the X-ray luminosity in the specific energy band where the data were taken. 

\citet{McCR2006} and \citet{Wu2010} compiled two samples of RXTE observations of Galactic NS and BH XRBs at different spectral states, for which they also calculated the best fit parameters of simple spectral models. Out of the two lists (which in some cases contained multiple observations of the same object at different spectral states) we chose all XRBs for which there was a distance measurement and a BH mass determination for BH XRBs. For the case of NS XRBs, we assumed a nominal mass of $1.4\, M_{\odot}$. Using the appropriate \emph{XSPEC} models, we calculated bolometric correction factors between the bolometric X-ray luminosity ($0.03-100\,\rm keV$) inferred by the best fitting model for each observation, and the X-ray luminosity in a specific energy band. We should note here that the fraction of the energy emitted by XRBs outside the $0.03-100\,\rm keV$ range is negligible, and hence the terms bolometric luminosity and X-ray bolometric luminosity can be used interchangeably. For each system we calculate these bolometric correction factors with and without absorption corrections. These factors would correspond to the observed and intrinsic X-ray luminosity at a specific energy band respectively.

According to \citet{McCR2006} and \citet{Wu2010} the XRBs in their compilations are in either a low-hard spectral state, where the spectrum is dominated by a power-law component, or in a high-soft state, where the spectrum is dominated by the thermal emission of the disk. \citet{McCR2006} also use a third category of a ``steep power-law'' state for BH XRBs with accretion rates close to the Eddington limit. These systems do show a soft spectrum, which however can be fit better with an absorbed steep power-law than with a thermal disk model. For our purposes, we do not treat separately systems in the high-soft and steep power law states, as they turn out to have very similar bolometric correction factors.

\begin{figure}
\centering
\includegraphics{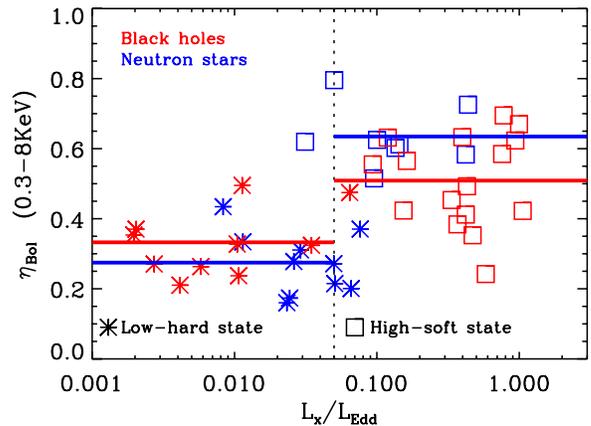}
\caption{\label{bol_cor} Bolometric correction factors for the $0.3-8\,\rm keV$ band, for each of the observations from the  \citet{McCR2006} and  \citet{Wu2010} samples. Star symbols denote systems in the low-hard state, while open squares systems in the high-soft state. Blue color corresponds to NS XRBs and red to BH XRBs. The dashed vertical line shows the approximate Eddington ratio at which the state transition happens. The horizontal lines denote the mean bolometric correction of NS or BH XRBs at the low-hard or high-soft state respectively.}
\end{figure}

In our calculations we treat separately NS and BH XRBs. The transition between low-hard and high-soft state in both cases happens at a luminosity of $\sim 5\%\, L_{Edd}$ \citep{McCR2006}. Figure~\ref{bol_cor} shows the correction factor ($\eta_{Bol}$) between the bolometric X-ray luminosity and the intrinsic (after correcting for absorption) luminosity in the $0.3-8\,\rm keV$ band (i.e. $\eta_{Bol}L_{Bol}=L_{0.3-8\,\rm keV}$), as a function of the Eddington ratio of the source. For each energy band we calculate the mean and the variance of the  bolometric correction for the high-soft and low-hard states, and BH and NS XRBs separately. Furthermore, we perform our calculations with and without correcting for absorption due to circumstellar or interstellar gas, corresponding to the observed and the intrinsic luminosity accordingly. The full list of the calculated bolometric corrections for different bands can be found in Table~\ref{bol_cor_table}. The mean values of the calculated bolometric corrections for the different energy bands are used in equations (1)-(6) of \citet{Fragos2008} in order to compare our PS models with observational data. The variances we calculate can enter in the estimation of the uncertainties of our models. In the rest of the paper, we use the absorption uncorrected values in our comparison, as these may be most easily compared with observed X-ray luminosities of galaxies. This is effectively equivalent with the assumption that all galaxies in the universe have similar intrinsic absorption as the Milky Way.

Although these bolometric corrections are based on spectral fits of RXTE data which have a low-energy limit of $\sim 2.5\rm\,keV$, the RXTE band covers most of the Wien tail of the disk black-body component, allowing a reliable measurement of the disk temperature \citep[e.g. Figures 4.8, 4.9, in ][]{McCR2006}. A comparison between the disk temperature based on RXTE measurements, with  similar measurements for the same objects in similar states derived from spectra extending to lower energies \citep[using data obtained with XMM-Newton, \emph{Suzaku}, Beppo-SAX, or ASCA; e.g.][]{Kotani2000,Kubota2005,Gou2009,Caballero2009,Tamura2012} shows very good agreement, indicating  that indeed the relatively restricted soft X-ray coverage of RXTE does not significantly bias the temperature of the thermal component of sources in the high-state.

\begin{deluxetable*}{lccccccccc}
%\rotate
\centering
\tablecolumns{10}
\tabletypesize{\scriptsize}
%\tablewidth{585pt}
\tablecaption{Mean value and standard deviation of the bolometric correction ($\eta_{bol}$) factor at different energy bands.   
\label{bol_cor_table}}
\tablehead{ 
   \colhead{} & 
   \multicolumn{4}{c}{Absorption Corrected} &
   \colhead{} & 
   \multicolumn{4}{c}{Absorption Uncorrected}\\
   \cline{2-5} \cline{7-10}   
   \colhead{} &
   \multicolumn{2}{c}{High-Soft State} &
   \multicolumn{2}{c}{Low-Hard State} &
   \colhead{} &
   \multicolumn{2}{c}{High-Soft State} &
   \multicolumn{2}{c}{Low-Hard State} \\
   \colhead{Energy band} &
   \colhead{NS} &
   \colhead{BH} &
   \colhead{NS} &
   \colhead{BH} &
   \colhead{} &
   \colhead{NS} &
   \colhead{BH} &
   \colhead{NS} &
   \colhead{BH} 
   }
\startdata
$0.3-2\,  \rm keV$ & 0.06$\pm$0.09 & 0.28$\pm$0.20 & 0.08$\pm$0.06 & 0.13$\pm$0.11 & & 0.05$\pm$0.08 & 0.09$\pm$0.10 & 0.05$\pm$0.03 & 0.08$\pm$0.06\\
$0.3-7\,  \rm keV$ & 0.61$\pm$0.09 & 0.75$\pm$0.18 & 0.29$\pm$0.12 & 0.36$\pm$0.13 & & 0.57$\pm$0.09 & 0.48$\pm$0.14 & 0.25$\pm$0.08 & 0.30$\pm$0.08\\
$0.3-8\,  \rm keV$ & 0.67$\pm$0.09 & 0.78$\pm$0.16 & 0.31$\pm$0.12 & 0.39$\pm$0.13 & & 0.63$\pm$0.09 & 0.51$\pm$0.13 & 0.27$\pm$0.09 & 0.33$\pm$0.10\\
$0.3-10\, \rm keV$ & 0.77$\pm$0.08 & 0.82$\pm$0.13 & 0.36$\pm$0.13 & 0.43$\pm$0.13 & & 0.73$\pm$0.08 & 0.55$\pm$0.12 & 0.32$\pm$0.10 & 0.38$\pm$0.10\\
$0.5-2\,  \rm keV$ & 0.06$\pm$0.09 & 0.26$\pm$0.19 & 0.08$\pm$0.06 & 0.12$\pm$0.09 & & 0.05$\pm$0.08 & 0.09$\pm$0.10 & 0.05$\pm$0.03 & 0.07$\pm$0.05\\
$0.5-10\, \rm keV$ & 0.77$\pm$0.08 & 0.80$\pm$0.11 & 0.36$\pm$0.13 & 0.42$\pm$0.13 & & 0.73$\pm$0.07 & 0.55$\pm$0.12 & 0.32$\pm$0.10 & 0.37$\pm$0.10\\
$2-10\, \rm keV$ & 0.71$\pm$0.08 & 0.54$\pm$0.12 & 0.28$\pm$0.09 & 0.30$\pm$0.11 & & 0.68$\pm$0.07 & 0.46$\pm$0.09 & 0.27$\pm$0.08 & 0.30$\pm$0.11
\enddata
\end{deluxetable*}

%%%%%%%%%%%%%%%%%%%%%%%%%%%%%%%%%%%%%%%%%%%%%%%%%%%%%%%%%%%%%%%%%%%%%%%%%%%%%
%%%%%%%%%%%%%%%%%%%%%%%%%%%%%%%%%%%%%%%%%%%%%%%%%%%%%%%%%%%%%%%%%%%%%%%%%%%%%
\section{Observational Constraints and Statistical Model Comparison}

\subsection{Available observational constraints}

To constrain our model parameters, we compare our models to observations of the X-ray properties of local galaxies. There have been several recent studies, where high quality multi-wavelength data are used to derive scaling relations between total X-ray emission from HMXBs and the galaxy-wide SFR  as well as a scaling between total X-ray emission from LMXBs and stellar mass, in the local universe. \citet{Boroson2011} selected a sample of 30 normal early-type galaxies, for all of which optical spectroscopy and Chandra data were available. From the optical data they estimated the stellar mass, while from spectral fitting of the X-ray data they estimated the contribution of unresolved LMXBs to the diffuse X-ray emission. We averaged the measurements for the \citet{Boroson2011} sample to calculate the X-ray emission from LMXBs per unit stellar mass, finding $log(L_{X,LMXBs\,[0.5-8\,\rm keV]}/M_{*})=29.18\pm 0.17\, \rm erg\,s^{-1}\, M_{\odot}^{-1}$. \citet{Lehmer2010} used Chandra observations of star-forming galaxies to study the galaxy-wide XRB emission and its correlation with SFR and stellar mass. They assumed that X-ray emission in the hard X-ray band ($2-10\, \rm keV$) of these galaxies can be decomposed into two components. The first component is related to HMXBs and should be proportional to the SFR of the galaxy, while the second component is related to LMXBs and should be proportional to the stellar mass of the galaxy. This simple model, which is described by the equation $L_{X\,[2-10\,\rm keV]}= \alpha M_{*}+ \beta SFR$, was fitted to their data, and the best fitting values were found to be: $log\alpha \equiv log(L_{X,LMXBs\,[2-10\,\rm keV]}/M_{*})=28.96\pm 0.34\,\rm erg\,s^{-1}\, M_{\odot}^{-1}$ and $log\beta \equiv log(L_{X,HMXBs\,[2-10\,\rm keV]}/SFR)=29.21\pm 0.34\,\rm erg\,s^{-1}\, M_{\odot}^{-1}\, yr$. Most recently, \citet{Mineo2012a}, based on a homogeneous set of X-ray, infrared and ultraviolet observations from the Chandra, Spitzer, GALEX and 2MASS archives, studied populations of HMXBs in a sample of 29 nearby star-forming galaxies and their relation to the SFR. In agreement with previous results, they found that HMXBs are a good tracer of the recent star-formation activity in the host galaxy and their collective luminosity scales with the SFR as $log(L_{X,HMXBs\,[0.3-8\,\rm keV]}/SFR)=29.47\pm 0.4\,\rm erg\,s^{-1}\, M_{\odot}^{-1}\, yr$.

\citet{TG2008} used multi-wavelength datasets of the CDFS, E-CDFS, CDFN, and XBootes fields to construct XLFs of normal galaxies in different redshift bins. The integral of these XLFs gives us the total X-ray luminosity per unit volume of normal galaxies in three redshift bins. \citet{TG2008} used the soft, $0.5-2\,\rm keV$, Chandra energy band, as Chandra is most sensitive in this band. However, this band likely contains also significant emission from hot diffuse gas. Hence we can only use these observed values as upper limits when comparing them to our simulations which do not include any emission from hot gas. We should note here, that in the last years there have been several X-ray stacking studies, especially in the Chandra deep fields \citep[e.g.][]{Lehmer2007,Lehmer2008,Symeonidis2011}, where the X-ray luminosity per unit  stellar mass or unit SFR, for early and late type galaxies respectively, has been measured at $z \approx 0-1.4$. However, a proper comparison of our models with these observational surveys would require a careful matching of the sample selection of these studies, which is outside the scope of this paper.

\subsection{Model likelihood calculation}

We compare the results by \citet{Boroson2011} and \citet{Lehmer2010} to the prediction of our models for $L_{X,\,LMXBs}/M_{*}$ at $z=0$ and in two different energy bands, and the results by \citet{Mineo2012a} and \citet{Lehmer2010} to our models' predicted $L_{X,HMXBs}/SFR$, again in two different energy bands. Furthermore, we compare our model predictions for the X-ray luminosity per unit volume with values reported by \citet{TG2008}, which we use as upper limits. The quantitative comparison is done by calculating the likelihood of the observations given a model, for each model, and then comparing the likelihood values.

\begin{deluxetable*}{ccccccccc} 
\centering
\tablecolumns{9}
\tabletypesize{\scriptsize}
\tablewidth{0pt}
\tablecaption{List of selected PS models that appear in the text and figures of this paper. A complete list of all models can be found in the online version of the article. 
\label{modellist}} 
\tablehead{ \colhead{Model} & 
     \colhead{$\alpha_{CE}$\tablenotemark{a}} & 
     \colhead{IMF exponent\tablenotemark{b}} & 
     \colhead{$\eta_{wind}$\tablenotemark{c}} & 
     \colhead{CE-HG\tablenotemark{d}} & 
     \colhead{q distribution\tablenotemark{e}} & 
     \colhead{$\kappa_{DC BH}$\tablenotemark{f}} &
     \colhead{Rank\tablenotemark{g}} &
     \colhead{$\log$(Likelihood ratio)\tablenotemark{h}} 
     }
\startdata
     245  &  0.1  &  -2.7  &  1.0  &  No  	  &  50-50  &  0.1     &     1    &         0.0000000\\  
     229  &  0.1  &  -2.7  &  2.0  &  Yes  	  &  50-50  &  0.0    &     2    &      -0.035048515\\   
     269  &  0.1  &  -2.7  &  1.0  &  Yes  	  &  50-50  &  0.1    &     3    &       -0.11226917\\   
     205  &  0.1  &  -2.7  &  2.0  &  No  	  &  50-50  &  0.0     &     4    &       -0.13235138\\  
     249  &  0.1  &  -2.35  &  2.0  &  No  	  &  50-50  &  0.1    &     5    &       -0.28666705\\   
     273  &  0.1  &  -2.35  &  2.0  &  Yes    &  50-50  &  0.1   &     6    &       -0.28926393\\
      53  &  0.1  &  -2.7  &  1.0  &  No  	  &  flat  &  0.1      &    20    &        -1.3945094\\
     246  &  0.2  &  -2.7  &  1.0  &  No  	  &  50-50  &  0.1     &    25    &        -2.0332112\\  
     241  &  0.1  &  -2.35  &  1.0  &  No  	  &  50-50  &  0.1    &    31    &        -2.3582808\\   
     253  &  0.1  &  -2.7  &  2.0  &  No  	  &  50-50  &  0.1     &    14    &       -0.90617906\\  
     261  &  0.1  &  -2.7  &  0.25  &  No  	  &  50-50  &  0.1    &    43    &        -3.4548391\\   
     197  &  0.1  &  -2.7  &  1.0  &  No  	  &  50-50  &  0.0     &    13    &       -0.80759106   
                                                                    
\enddata

\tablenotetext{a}{CE efficiency parameter}
\tablenotetext{b}{Exponent of the high-mass power law component of the IMF: \citet{Kroupa2001} (-2.35) or \citet{KW2003} (-2.7).}
\tablenotetext{c}{Stellar wind strength parameter with which the ``standard'' \citep{Belczynski2010} stellar wind prescription is multiplied.}
\tablenotetext{d}{Yes: all possible outcomes of a CE event with a HG donor are allowed, No: A CE with a HG donor star will always result to a merger.}
\tablenotetext{e}{Binary mass ratio distribution. ``50-50'' indicates half of the binaries originate from a ``twin binary'' distribution and half from flat mass ratio distribution.} 
\tablenotetext{f}{Parameter with which the ``standard'' \citet{Hobbs2005} kick distribution is multiplied for BHs formed though a SN explosion with negligible ejected mass.}
\tablenotetext{g}{ Ranking of the model in our statistical comparison.}
\tablenotetext{h}{ Ratio of the likelihood of the observations given a model to our maximum likelihood model ($\log{L(O|M_i)}-\log{L(O|M_{ref})}$).}

\end{deluxetable*}

\begin{figure*}
\centering
\includegraphics{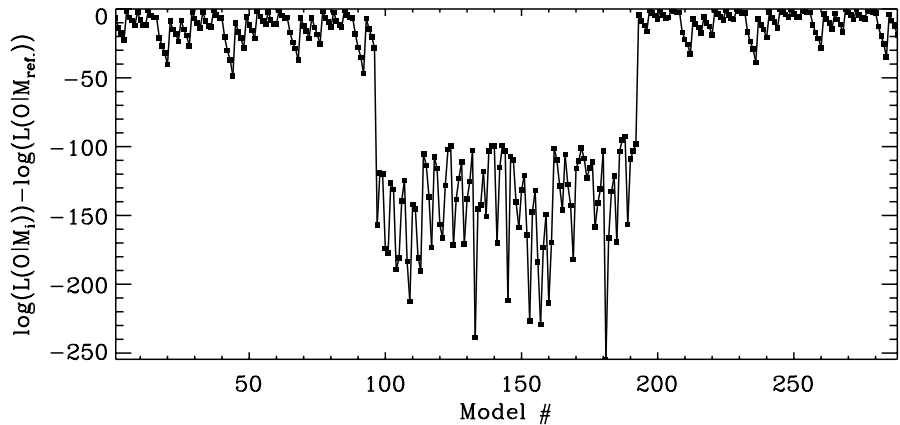}
\caption{\label{likelihood} Ratio of the likelihood of the observations given a model to our maximum likelihood model.}
\end{figure*}

Assuming that the reported errors for all observed quantities are gaussian in the log space, then the likelihood of the data given a model becomes: 
\begin{equation} L_i(D|M,{f_j}) = \prod\limits_{j = 1,7} {G({M_{i,j}};{f_j}{\mu _j},{\sigma _j})} P({f_j}) \end{equation} 
where $G$ is the Gaussian function, $\mu_j$ and $\sigma_j$ for $j=1,4$ are the observed values and the associated errors at $z=0$ for $log(L_{X,LMXBs\,[0.5-8\,\rm keV]}/M_{*})$ \citep{Boroson2011}, $log(L_{X,LMXBs\,[2-10\,\rm keV]}/M_{*})$ \citep{Lehmer2010}, $log(L_{X,HMXBs\,[2-10\,\rm keV]}/SFR)$ \citep{Lehmer2010}, and $log(L_{X,HMXBs\,[0.3-8\,\rm keV]}/SFR)$ \citep{Mineo2012a}, $\mu_j$ and $\sigma_j$ for $j=5,7$ are the observed values and the associated errors at $z\sim 0.13$, $z\sim 0.38$, and $z\sim 0.78$ for $log(L_{X\,[0.5-2\,\rm keV]}/Volume)$ \citep{TG2008}. $M_{i,j}$ are the predictions of the $i^{th}$ model for each of these quantities and $f_j$ is the fraction of the luminosity density $\mu_j$ assumed to be due to X-ray binaries. $P(f_j)$ is the prior probability for a given value of $f_j$. Since we are not attempting to constrain the X-ray binary fraction $f$ we marginalize over $f$, i.e., \begin{equation} L_{i,j}({M_{i,j}}|D) = \int {{L_i}(D|{M_{i,j}},{f_j})P(} {f_j})d{f_j} \end{equation} For the $z=0$ points we assume that the luminosity densities are purely due to X-ray binares and therefore $f$=1. In this case $P(f_{j=1,4}) = \delta(f_j=1)$. We assume a flat prior for the X-ray binary fraction for the $z>0$ luminosity densities, i.e., $P(f_j)=C_j$, where the constants $C_j$ are chosen so that the likelihood function remains normalized to unity. Noting that: 
\begin{equation} 
\int_0^1 {G({M_{i,j}};{f_j}{\mu _j},{\sigma _j}){C_j}d{f_j}} = \int_0^{{\mu _j}} {G({M_{i,j}};{{\mu^{\prime}}_j},{\sigma _j}){C_j}d} {\mu^{\prime}_j} =  
\end{equation} 
$$=\left| {erf\left[ {\frac{{{{\left( {{\mu _j} - {M_{i,j}}} \right)}^2}}}{{2\sigma _j^2}}} \right] - erf\left[ {\frac{{{M^2}_{i,j}}}{{2\sigma _j^2}}} \right]} \right| $$
	The marginalized likelihood is then 
\begin{equation} 
{{\rm{L}}_{{\rm{i,j}}}}{\rm{(}}{{\rm{M}}_{{\rm{i,j}}}}{\rm{|D)}} = 
\end{equation} 
$$=\prod\limits_{j = 1,4} {G({M_{i,j}};{\mu _j},{\sigma _j})} \times \prod\limits_{j = 5,7} {\left| {erf\left[ {\frac{{{{\left( {{\mu _j} - {M_{i,j}}} \right)}^2}}}{{2\sigma _j^2}}} \right] - erf\left[ {\frac{{{M^2}_{i,j}}}{{2\sigma _j^2}}} \right]} \right|} $$
	
Figure~\ref{likelihood} shows the likelihood ratio of each of our 288 models to the maximum likelihood model ($L(D|M_{ref.})$). This effectively means that we renormalized the likelihood of the observations given a model, so that our maximum likelihood model has a likelihood value of 1. A list of the likelihood values of each model can be found in Table~\ref{modellist}. One should note that the scale of the Y-axis in Figure~\ref{likelihood} is logarithmic, and only a small fraction of all the 288 models we considered have a likelihood close to the our reference, maximum likelihood model (Models~245). Namely, only 14 out of 288 models have likelihood values within a factor of 10 from the maximum likelihood model, and only 24 within a factor of 100 respectively, which is an encouraging evidence that our calculated likelihood is a good discriminator between the different models. One other apparent feature of Figure~\ref{likelihood} is that models Models 100-199 have systematically lower likelihood values than the rest. In all these models the initial binary mass ratio is assumed to be ``twin'', i.e. a flat distribution between 0.9 and 1.0, which always results in an LMXB population inconsistent with observations (see also Section 5.1 for further discussion).

\begin{figure*}
\centering
\includegraphics{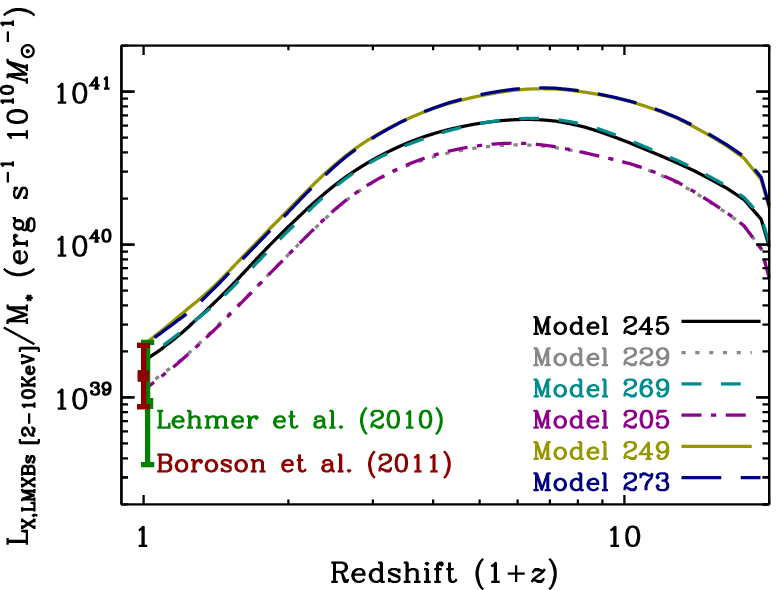}
\includegraphics{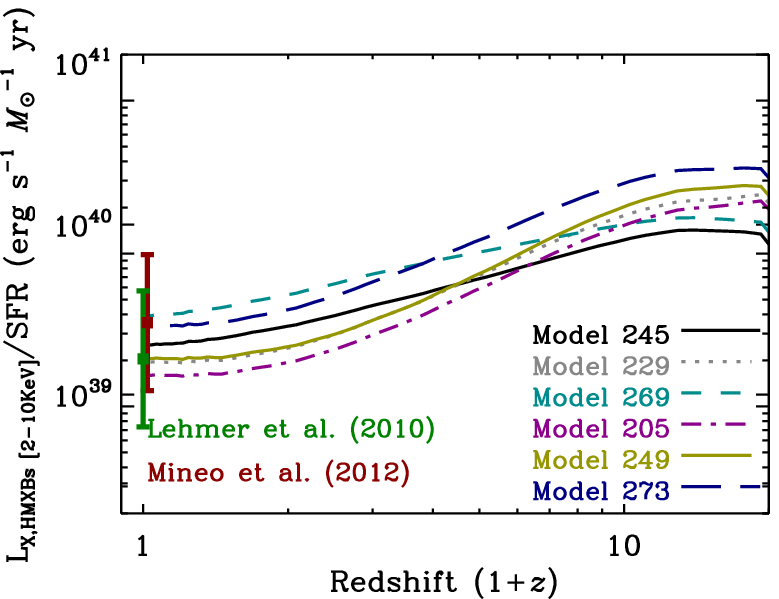}\\
\includegraphics{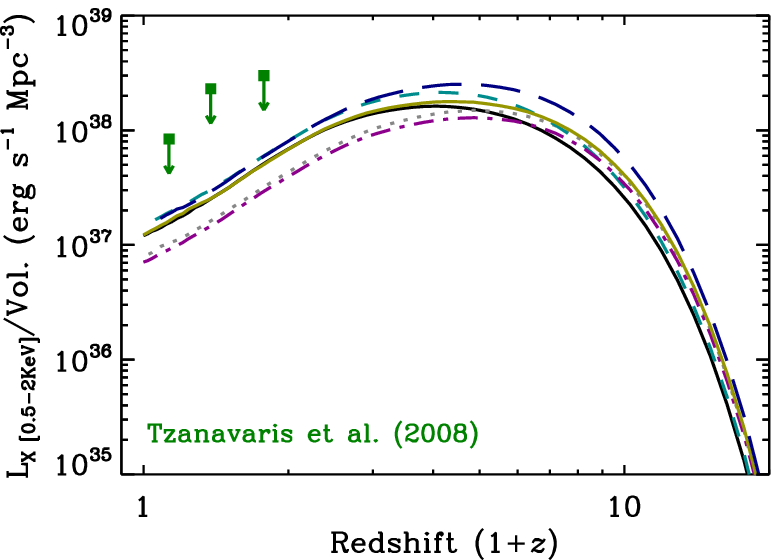}
\caption{\label{Model_constraints} X-ray luminosity in the $2-10\,\rm keV$ range, from the LMXB population, per unit stellar mass (top left panel) and X-ray luminosity in the same range, from the HMXB population, per unit SFR (top right panel), as a function of redshift, for the six models that satisfy within one $\sigma$ all the observational constraints.The green squares show the observed values for  $L_{X,LMXBs [2-10\,{\rm keV}]}/M_{*}$ (top left panel) and $L_{X,HMXBs [0.5-8\,{\rm keV}]}/SFR$ (top right panel) as derived by \citet{Lehmer2010}. The red squares show the observed constraints for $L_{X,LMXBs [0.3-8\,{\rm keV}]}/M_{*}$ by \citet{Boroson2011}, and for $L_{X,HMXBs [0.5-8\,{\rm keV}]}/SFR$ by \citet{Mineo2012a} after converted for demonstration purposes to the $2-10\,\rm keV$ energy band, in the left and right panel respectively. The bottom panel shows the X-ray luminosity density coming from the the whole XRB population as a function of redshift. The green squares show the observed values as reported by \citet{TG2008}, which we used as upper limits in our statistical analysis. The parameters of the different models are shown in Table~\ref{modellist}}
\end{figure*}

Figure~\ref{Model_constraints} shows the X-ray luminosity in the $2-10\,\rm keV$ range, from the LMXB population, per unit stellar mass (top left panel) and X-ray luminosity in the same range, from the HMXB population, per unit SFR (top right panel) as a function of redshift. The six models shown in this figure are the ones that satisfy within one $\sigma$ all the observational constraints, which are also the six models with the highest likelihood values. The observed values of these two quantities for $z=0$ by \citet{Lehmer2010}, and by \citet{Boroson2011} and \citet{Mineo2012a} after converting them to the $2-10\,\rm keV$ range, are also shown for comparison. The bottom panel of Figure~\ref{Model_constraints} shows the X-ray luminosity density coming from the the whole XRB population as a function of redshift. The observational constraints derived by \citet{TG2008}, shown as green squares in the figure, are used as upper limits in our statistical analysis. Figure~\ref{Model_constraints} demonstrates the robustness of our model predictions. All six maximum likelihood models exhibit a very similar behavior for the whole range of redshifts, and the differences between the different models at higher redshifts are equivalent to the uncertainties of the observed values in the local universe.   

\subsection{The specific X-ray luminosity of local elliptical galaxies}

\citet{KF2010}, using a sample of local elliptical galaxies with measured stellar ages, found that the fraction of bright LMXBs in ``young'' elliptical galaxies is higher compared to ``old'' ellipticals.  More recently, \citet{ZGA2012}, using a similar sample of nearby ellipticals, reported a positive correlation between the normalization of the XLFs of these galaxies, and the luminosity average age of their stellar population, as estimated by single stellar population spectral energy distribution fitting results. The specific X-ray luminosity in these galaxies ($L_X/M_*$), which depends both on the shape and the normalization of the XLF, shows little to no evolution with the estimated stellar age of the galaxies \citep[see also ][]{Boroson2011}, which at face value seems to be inconsistent with our prediction for the evolution of $L_X/M_*$ with redshift. However, the galaxy sample in both of these studies shows also a strong correlation of the estimated stellar age with the globular cluster specific frequency ($S_{N}$), as typical nearby old giant ellipticals tend to have a high $S_{N}$. At the same time a strong correlation between  $L_{X}/M_{*}$ and $S_{N}$ has been observed in these galaxies \citep{Boroson2011}. In addition, the age estimates of the observed galaxies are derived under the assumption of a single age stellar population, and assigning single ages to ``young'' ellipticals will suffer from large uncertainties in the stellar ages due to very young populations mixing with old stellar populations \citep[see, e.g.,][]{Idiart2007}. Hence, it is hard to disentangle whether the observed correlation reported by  \citet{ZGA2012} means that the specific X-ray luminosity of the LMXB population in these galaxies indeed does not evolve with stellar population age, or it is an effect of the underlying correlations in the properties of the galaxy sample they used, and uncertainties in the age estimates. In either case, we should stress again here that our models consider only field XRBs and a general galaxy population, where the dynamically formed XRBs are expected to have minimal contribution. In contrast, the observed correlations discussed above are driven by globular cluster rich old elliptical galaxies whose LMXB population is probably dominated by dynamically formed LMXBs \citep[e.g.,][]{Irwin2005}.

\begin{figure}
\centering
\includegraphics{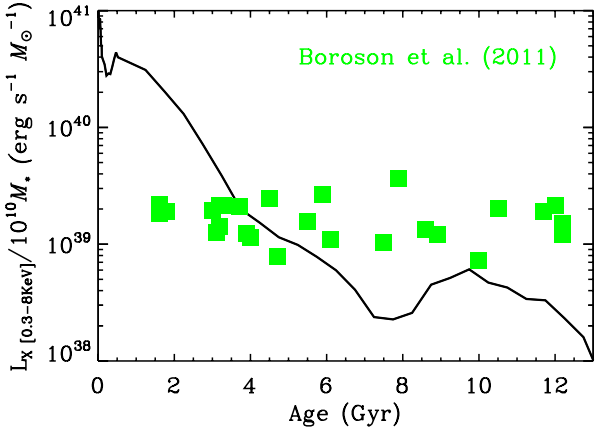}
\caption{\label{ellipticals}Specific  X-ray luminosity in the  $0.3-8\,\rm keV$ energy range, i.e. luminosity per unit stellar mass, of a coeval stellar population as a function of the population's age for our ``reference'' model (Model 245; see also sections 2.4 and 4.2) and for solar metallicity. On the same figure over-plotted are the specific X-ray luminosities, in the same energy band, of nearby elliptical galaxies that have available in the literature estimates of their stellar population age. These data points are derived from the galaxy sample presented in \citet{Boroson2011}.}
\end{figure}

Figure~\ref{ellipticals} shows the evolution of the specific X-ray luminosity from a coeval stellar population (similar to Figure~\ref{single_age}, but in the $0.3-8\,\rm keV$ energy range) as a function of the stellar population age for our ``reference'' model (Model 245; see also sections 2.4 and 4.2) and for solar metallicity. In the same figure we have over-plotted the $L_X/M_*$ in the same energy range for elliptical galaxies with stellar age estimates from the \citet{Boroson2011} sample. The galaxy sample in \citet{Boroson2011} is very similar to the sample by \citet{ZGA2012}.  We want to stress here that a direct comparison of the model curve to the observed data points is not appropriate for two main reasons: the model curve neglects the dynamically formed LMXB population which can be significant for some of these galaxies, and the age estimates of the observed galaxies are derived under the assumption of a single age stellar population. However, assigning single ages to young ellipticals will suffer from large uncertainties in the stellar ages due to very ``young'' populations mixing with old stellar populations. Keeping in mind these two important caveats, we can make a few instructive observations: \emph{(i)} Comparing the model curve to the observed $L_X/M_*$ of the ``old'' ($\sim 9-12 \,\rm Gyr$) ellipticals, we find that our models  under-predict the LMXB specific luminosity. This is to be expected, as in these galaxies the dynamically formed LMXB population is significant, and our models consider only the field LMXB population. \emph{(ii)} Comparing the model curve to the ellipticals with ages $\sim 4-5 \,\rm Gyr$ (which also tend to have lower specific globular cluster frequencies), we find that our models predict the same $L_X/M_*$ as the observations. This can be also seen in the top left panel of Figure~\ref{Model_constraints}. The average mass-weighted age of the stellar population at $z=0$ based on the Millennium II simulation is $\sim 4.7 \,\rm Gyr$, and the predicted from our models $L_X/M_*$ (coming from LMXBs) is in very good agreement with the $L_X/M_*$ as reported by \citet{Boroson2011}. \emph{(iii)} Our model curve is in disagreement with the observationally determined $L_X/M_*$ of the ``young'' ellipticals (ages of $\sim 1-2 \,\rm Gyr$). However, these are the galaxies for which the single age stellar population assumption most likely breaks down. This makes the ages highly uncertain, as a small fraction of young stars ($<10\%$) can make a galaxy with a much older population appear only a few Gyr old \citep[see, e.g.,][for a discussion on this]{Idiart2007}. 

In conclusion, we believe that further work will be needed, both observational in order to derive more reliable star-formation history information and theoretical incorporating dynamically formed LMXBs in our models,  to unambiguously determine whether the sharp increase in $L_X/M_*$ going to ages younger than $\sim 3 \,\rm Gyr$ is real. This is clearly outside the scope of this study. However, the success of our models in predicting local $L_X/M_*$ and $L_X/SFR$ relations and the evolution of $L_X/SFR$ \citep{BasuZych2012} while be consistent at the same time with X-ray stacking results of the evolution of $L_X/M_*$ from early-type galaxies \citep{Lehmer2007} suggests that we are on the right track.

%%%%%%%%%%%%%%%%%%%%%%%%%%%%%%%%%%%%%%%%%%%%%%%%%%%%%%%%%%%%%%%%%%%%%%%%%%%%%
%%%%%%%%%%%%%%%%%%%%%%%%%%%%%%%%%%%%%%%%%%%%%%%%%%%%%%%%%%%%%%%%%%%%%%%%%%%%%
\section{Results}

\subsection{Effects of the different model parameters on the X-ray binary population}

As already mentioned, our models have two types of parameters, the ones that correspond to distributions of initial properties of the stellar population, and which are guided by observations, and the ``true'' free parameters, that correspond to poorly understood physical processes which we are not able to model in detail. For both types of parameters, it is instructive to study the effect they have in the XRB population and the constraints we can put on the values of these parameters by comparing the predictions of our models with observations. For the first type of parameters, this comparison allows us to infer which part of the observationally allowed parameter space is favored by our models. Constraining the second group of parameters is more essential, as we improve our understanding of complex physical processes that we cannot model in detail.

\begin{figure*}
\centering
\includegraphics{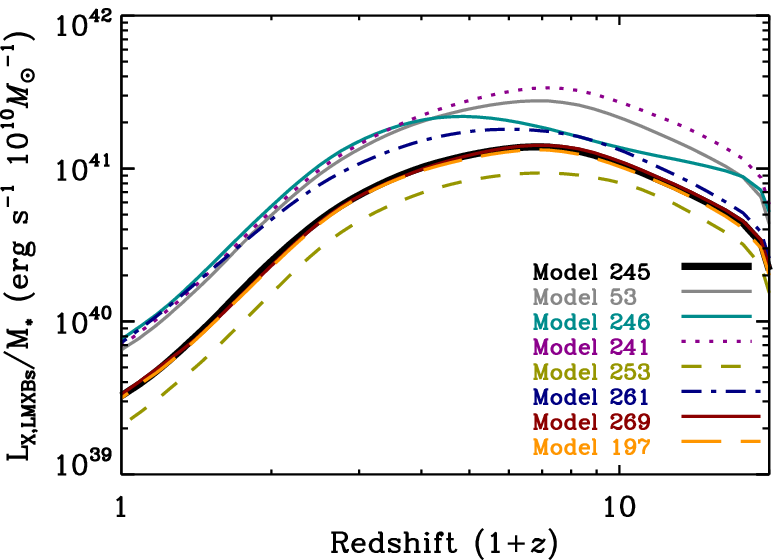}
\includegraphics{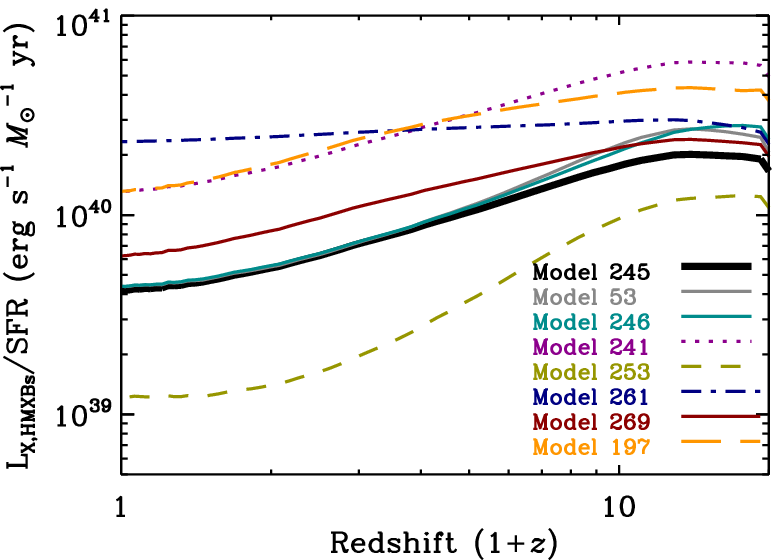}
\caption{\label{Model_behavior_param} Bolometric X-ray luminosity of the LMXB population, per unit stellar mass (left panel) and X-ray luminosity of the HMXB population, per unit SFR (right panel), as a function of redshift, for eight selected models that show the dependence of the models' predictions to the different model parameters. Our maximum likelihood model (Model 245) is shown with the bold solid (black) line. The parameters of the different models are shown in Table~\ref{modellist}}
\end{figure*}

Figure~\ref{Model_behavior_param} shows the X-ray luminosity of the LMXB population, per unit stellar mass (left panel) and the X-ray luminosity of the HMXB population, per unit SFR (right panel), as a function of redshift, for eight selected models. The models shown in this figure were selected so that they demonstrate how each of the model parameters we varied in our grid affects the LMXB and HMXB population respectively. 

The CE efficiency $\alpha_{ce}$, i.e. the efficiency with which orbital energy is converted into thermal energy that will expel the donor star's envelope during the CE phase, is a parameter that mainly affects the formation rate of LMXBs. Most LMXBs formed in the field, through the evolution of primordial binaries, have to go through a CE phase in order to form a tight enough binary that will later on become a LMXB. A higher $\alpha_{ce}$ would result in more systems surviving the CE and forming tight post-CE binaries, and hence a higher formation rate of LMXBs. On the other hand, HMXBs do not necessarily have to go through a CE, but can instead become bright HMXBs through alternative formation channels \citep[e.g.][]{Valsecchi2010}. \emph{Model~246} has the same model parameters as our maximum likelihood model (\emph{Model~245}) except for the higher CE efficiency. Comparing the two models (see Figure~\ref{Model_behavior_param}), we observe that indeed the LMXB population is heavily affected by a small change in $\alpha_{ce}$, while the HMXB population is rather insensitive. Our parameter study shows that only a low CE efficiency ($\alpha_{ce}\sim 0.1$) produces XRB populations consistent with observations. This finding may alternatively be interpreted in terms of the binding energy parameter $\lambda$. Instead of the assumed value ($\lambda=1$) this parameter may be much lower ($lambda \sim 0.1$, if $\alpha_{\rm ce}$ is assumed 1) indicating that at the onset of the CE phase LMXB progenitors have highly bound envelopes. This in fact seems to be true for massive stars based on the actual estimates of envelope binding energies \citep[see Fig.1-4 of][]{Dominik2012}. 

Changing the initial binary mass ratio distribution between a flat distribution and 50\% flat - 50\% twin distribution affects the XRB population in a similar way (compare Model~53 and Model~245 in Figure~\ref{Model_behavior_param}). The progenitor of a LMXB is a binary with high mass ratio. By forcing half of the systems to have an initial mass ratio close to 1, the formation rate of LMXBs is decreased by a factor of two. On the other hand, the HMXB population is largely unaffected. This suggests that in a binary population with a flat initial mass ratio distribution, approximately 10\% of HMXBs have as progenitors primordial binaries with a mass ratio between 0.9 and 1. The statistical comparison of our models with observations favors a mixed initial mass ratio distribution of 50\% flat - 50\% twin distribution, which is consistent with observational surveys of young binaries \citep[e.g.][]{PS2006}, although some recent studies hint towards a single flat distribution \citep{SE2011}. A pure twin mass ratio distribution, although it is included in our parameter study (Models 100-199), results in models that are inconsistent with observations, as it will completely prevent the formation of LMXBs. Figure~\ref{likelihood} shows that models with a pure twin mass ratio distribution (Models 100-199) have systematically lower likelihood values than the rest.

Stellar winds play also an important role in the formation and evolution of XRBs, and they do so primarily in two ways.  A stronger stellar wind will increase the accretion rate on the compact object of a wind-fed HMXB, making it more luminous. On the contrary, a weaker stellar wind will result in smaller overall mass loss of the primary star that will form the compact object, and hence in the formation of more numerous and more massive BHs. BH-XRBs tend to be more luminous than XRBs with NS accretors, as on the one hand they can form stable RLO XRB systems with massive companion stars, and on the other hand they can drive in general higher accretion rates (compared again to NS accretors), both in RLO and wind-fed systems, because of the higher mass of the BH accretor. Furthermore, a reduced stellar wind would result in less orbital expansion due to angular momentum loss from the wind, and in overall more systems that will encounter Roche-lobe overflow mass transfer. It turns out that the latter two effects are the dominant ones. Comparing \emph{Model~253} ($\eta_{Wind}=2.0$) and \emph{Model~261} ($\eta_{Wind}=0.25$) to our maximum likelihood model (Model 245, $\eta_{Wind}=1.0$), one can see the large effects:  for stronger winds (model 253) the LMXB emission and HMXB emission is suppressed and for weaker winds (model 261) both the LMXB and HMXB emission is boosted. Our statistical analysis showed that the ``standard'' wind prescription ($\eta_{Wind}=1.0$), compiled and calibrated by \citet{Belczynski2010}, is favored, but models with stellar winds increased by a factor of two are also consistent with observations. If this result is taken at the face value (i.e., if potential degeneracies with other free parameters are neglected) it indicates that there is not much more unaccounted clumping in winds of massive stars that in the past led to overestimates of mass loss rates.

The power law slope at the high-mass end of the IMF affects the population of XRBs in a way equivalent to stellar winds, in the sense that a flatter IMF will produce relatively more BHs compared to a steeper one. Hence, a flatter IMF will result in more luminous populations of both LMXBs and HMXBs (compare \emph{Model~241} to \emph{Model~245}). In our standard treatment of supernova events, the most massive BHs receive no kick during their formation. Allowing for a small kick (10\% of the standard kicks that NSs receive) in these systems does not affect significantly the LMXB population, but results in an overall less luminous HMXB population (compare Model 245 to Model 197). HMXBs originate from more massive binaries on average, where the formation of a BH through direct core-collapse is more likely. Thus, allowing for small kicks in these supernova events leads to the disruption of binaries that would otherwise form and HMXB.

Finally, we found that allowing or not for all possible outcomes in CE phases with donor stars in the Hertzsprung gap, although it affects the shape of the XLF for a given galaxy \citep{Luo2012}, has a negligible effect on the integrated X-ray luminosity of a LMXB population and slightly increases the luminosity of a HMXB population (compare Model 245  to Model 269). This comes from the fact that LMXB progenitors are on average less massive than stars producing HMXBs. The probability of a CE phase to happen while the donor star is on the Hertzsprung gap is very low, since only very massive stars experience  significant expansion during Hertzsprung gap. In contrast, for massive binaries the Hertzsprung gap CE happens frequently and if the survival is allowed some systems survive and form extra population of HMXBs. This additional sources lead to a factor of  $\sim 1.5$ increase in the integrated HMXB X-ray luminosity throughout the cosmic time. This is very similar effect as noted for double compact object populations, for which massive BH-BH progenitors are greatly affected by Hertzsprung gap CE while lighter NS-NS progenitors are not \citep{Belczynski2007}.

\subsection{Evolution of global scaling relations with redshift}

\begin{figure}
\centering
\includegraphics{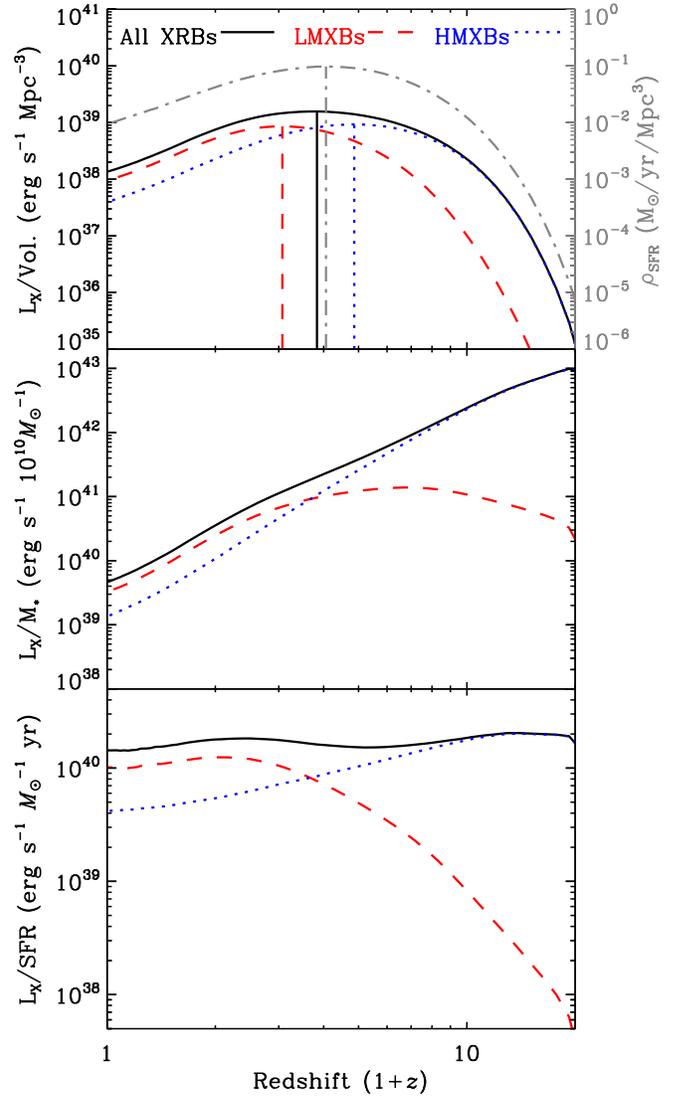}
\caption{\label{Model_all} Bolometric X-ray luminosity per unit volume (top panel), unit of stellar mass (middle panel), and unit of SFR (bottom panel), as a function of redshift, for our maximum likelihood model (Model~245). The solid (black) lines correspond to the whole XRB population, while the dotted (blue) lines show the contribution of the HMXBs, and the dashed (red) lines the contribution of the LMXBs. For comparison, the SFR density is shown on the top panel with a grey dash-dotted line. The vertical solid (black), dotted (blue), dashed (red), and dash-dotted (grey) lines denote the redshifts at which the peaks of the X-ray luminosity density from the whole XRB population, the HMXBs, and the LMXBs, and the SFR density occur.}
\end{figure}

Figure~\ref{Model_all} shows the bolometric X-ray luminosity per unit volume, unit of stellar mass, and unit of SFR, as a function of redshift, for our maximum likelihood model (Model~245). The relative contribution in these quantities of the LMXB and HMXB populations is also shown. Comparing in the top panel of Figure~\ref{Model_all}  the redshift evolution of the SFR density to the evolution of the X-ray luminosity density coming from LMXBs, one notices an evident time difference between the peaks of the two quantities. The peak of the SFR density occurs at  $z \sim 3.1$, while the peak of the X-ray luminosity density coming from the LMXB populations happens at $z\sim 2.1$. This delay between the two peaks translates into a typical delay time of $\sim 1.1\rm \, Gyr$ between the formation of a progenitor binary star and time that the systems reaches the LMXB phase. In earlier work by  \citet{WG1998,GW2001}, this timescale was a free parameter in their models which they we not able to constrain. We should note here, that the value of timescale is dependent on the exact definition of what one calls a LMXB. As mentioned earlier, in this work we define as LMXB any RLO XRB with a donor star less massive than $3\,\rm M_{\odot}$. Shifting this mass boundary to lower masses, will lengthen the delay.

Comparing now the SFR density peak to the peak of the X-ray luminosity density coming from HMXBs, one would expect that two peaks occur simultaneously. However, the peak X-ray luminosity density coming from HMXBs occurs at $z\sim 3.9$, $\sim 0.5\rm \, Gyr$ before the peak of the star formation. As we can see from the bottom panel of Figure~\ref{Model_all}, the X-ray luminosity coming from HMXBs, per unit SFR, is not constant with time, instead it is increasing with redshift. This is a result of the metallicity evolution. The mean metallicity of the newly formed stars is decreasing with redshift (see Figure~\ref{MRII} lower panel). A lower metallicity translates into weaker stellar winds, which in turn results in more numerous and more massive BHs. Since BH-HMXBs tend to be more luminous than NS-HMXBs, the  X-ray luminosity coming from HMXBs per unit SFR is higher for lower metallicity stellar populations, and hence at higher redshifts. Coming back to the time difference between the peak of the SFR and the peak of the X-ray luminosity density from HMXBs, it is now evident that it is because of the dependence of the HMXB population to the metallicity and the evolution of the metallicity with redshift.

One additional point that is demonstrated in Figure~\ref{Model_all} is that LMXBs dominate the X-ray emission from XRBs in our Universe today by a factor of $\sim 2.5$. The transition between an HMXB and LMXB dominated XRB population happens at a $z \sim 2.5$. This finding implies that the X-ray luminosity of an XRB population in a ``typical'' galaxy, with a star-formation history similar to the global star-formation history of the Universe, is dominated by the X-ray emission of LMXBs. This is not an unexpected result, given that in the Milky Way, which is a typical spiral galaxy, there are only a few HMXBs with luminosity above $10^{37}\rm\, erg\,s^{-1}$ \citep{GGS2002,VA2010}. However, our finding should serve as an extra caution to a usual misconception that the XRB population of galaxies that currently show some  star formation have a negligible LMXB population. \citet{Lehmer2010} showed observational evidence that the X-ray luminosity from galaxies with SFR as high as $\sim 2 \, \rm M_{\odot}\, yr^{-1}$ has a non-negligible contribution from LMXBs. 

Our modeling predicts that the X-ray luminosity coming from HMXBs, per unit SFR is increasing with redshift, and that this is a result of the metallicity evolution of the Universe. The evolution of the X-ray luminosity from HMXBs per unit SFR predicted by our models (by a factor of $\sim 10$ out to $z\sim 15$) is consistent with constraint set by the observed cosmic X-ray background \citep{Dijkstra2012}. Although the X-ray luminosity coming from HMXBs, per unit SFR, changes by a factor of $~5$ between $z=0$ and $z\sim 10$, the observed signature of this effect would be masked by the contamination of LMXBs. The bottom panel of Figure~\ref{Model_all} shows that the contamination of LMXBs operates in such a way that the X-ray luminosity from the whole XRB population, per unit SFR, remains approximately constant with time. These results are in agreement with the findings by \citet{Lehmer2008} and \citet{Mineo2012b} for the late-type galaxy population. Hence, our modeling suggests that the effect of the metallicity evolution to the HMXB population can only be observed with a carefully selected galaxy sample, where all galaxies have high enough SFR to stellar mass ratios, ensuring that their XRB populations are dominated by HMXBs.       

The middle panel of Figure 7 shows that the X-ray luminosity of the whole XRB population, per unit stellar mass, evolves with redshift following approximately a single power law from $z=0$, back to  $z \sim 15$. This relation probes two distinct phases of the cosmic evolution of the XRB population. At $z\lesssim 2.5$, where the emission from LMXBs dominates the total X-ray luminosity, the evolution of the X-ray luminosity per unit stellar mass is related to the average age of the LMXB population. The fact that relatively younger stellar populations are associated with more luminous populations of LMXBs has been already observed in X-ray stacking studies of distant galaxies \citep{Lehmer2007}. \citet{Lehmer2007} investigated the  average X-ray properties of early-type galaxies within the Extended Chandra Deep Field-South, and found a slight increase of $L_{X}/L_{B,\odot}$, where $L_{B,\odot}$ is the optical luminosity in the B-band, with redshift, over the redshift range $z=0-0.7$, in excellent agreement with our model predictions. At  $z\gtrsim 2.5$, it is the emission from HMXBs that dominates the X-ray luminosity of the XRB population. The total X-ray luminosity coming from HMXBs is to first order, neglecting the metallicity evolution effects, proportional to the SFR. Hence, the evolution of X-ray luminosity per unit stellar mass at  $z\gtrsim 2.5$, can be a probe of the evolution of the specific SFR (SFR per unit stellar mass) with redshift.

In the interpretation of our simulation results, and the comparison with the observational data, one should keep in mind that all the quantities calculated in this paper correspond to the average of all galaxies in the Universe. A more detailed comparison of our models with galaxy surveys of early and late type galaxies separately, where all the sample selection effects are taken into account, is discussed in other works \citep[][; Hornschemeier et al. 2012, in prep.]{Tremmel2012,BasuZych2012}.

%%%%%%%%%%%%%%%%%%%%%%%%%%%%%%%%%%%%%%%%%%%%%%%%%%%%%%%%%%%%%%%%%%%%%%%%%%%%%
%%%%%%%%%%%%%%%%%%%%%%%%%%%%%%%%%%%%%%%%%%%%%%%%%%%%%%%%%%%%%%%%%%%%%%%%%%%%%
\section{Summary and Conclusions}

We present here the first PS study of XRBs in a cosmological context. We created the largest grid of 288 PS models over nine metallicities, where we varied all the major model parameters that are known to affect the evolution of XRBs. We convolved these models with information about the star-formation history and metallicity evolution of the Universe from the Millennium II simulation and the \citet{Guo2011} semi-analytic galaxy catalog. The combination of our PS models with a cosmological simulation allowed us to derive global scaling relations of the XRB population with properties such as SFR and stellar mass, and the evolution of these relations with redshift. These predictions were compared statistically with observations from the local Universe in order to constrain our models and make robust predictions about the high-redshift Universe. The main conclusions of this work can be summarized as follows:

\begin{enumerate}

\item The statistical comparison of our model predictions revealed several models that are in excellent agreement with observations of the local Universe. Furthermore, we found that all models that are consistent with observational data in the local Universe exhibit very similar behavior at higher redshifts, indicating that our predictions for the redshift evolution of XRB populations are robust. 

\item Although the main goal of this paper is not to constrain the parameters of our PS models but instead to study the redshift evolution of XRB populations, our parameter study allows for some general conclusions on the possible values of the PS parameters we varied. The statistical analysis showed that a CE efficiency $\lambda\times\alpha_{CE}\lesssim 0.1$, where $\lambda$ is a measure of the central concentration of the donor star, and possibly a mixed initial mass ratio distribution (i.e. 50\% of binaries are drawn from a flat initial mass ratio distribution and 50\% from a ``'twin'' distributuon) is necessary in order to produce a LMXB population consistent with observations. These two parameters show some degeneracy, and a slightly lower CE efficiency would possibly favor a flat initial mass ratio distribution. On the other hand, model parameters that regulate the formation rate of BH-XRBs, such as the IMF and the stellar wind strength, appear to be important for both the LMXB and HMXB populations. This is not surprising as XRBs with BH accretors can be very luminous, and even a relatively small number of them can dominate the integrated X-ray luminosity of a population. Again, these parameters show a degeneracy as they affect the population in a similar way, and only their combined effect can be constrained from observations. 

\item The X-ray luminosity of XRBs per unit stellar mass or unit star-formation rate, is  significantly higher for low-metallicity stellar populations, due to the fact that the young massive O/B stars, which are the precursors to compact object accretors in X-ray binaries, will lose less stellar mass from line-driven winds over their lifetimes for lower metallicities. This results in more numerous and more massive BHs, and subsequently more luminous X-ray binary populations.

\item The X-ray emission from XRBs in the local Universe is dominated by LMXBs, and it is only at $z \gtrsim 2.5$ that HMXBs start to dominate the X-ray emission. The redshift that this transition happens is lower than that of the peak of star formation ($z\sim 3$), which shows that there is a delay between the formation of primordial binaries and the time at which these binaries become LMXBs. The formation timescale of LMXBs is estimated to be $\sim 1.1\,\rm Gyr$. 
This finding also implies that the X-ray luminosity of an XRB population in a ``typical'' galaxy, with a star-formation history similar to the global star-formation history of the Universe, is dominated by the X-ray emission of LMXBs.

\item The formation rate of HMXBs follows closely the SFR. However, the X-ray luminosity from HMXBs per unit SFR ($L_{X,\, HMXBs}/SFR$) depends also on metallicity.  Taking into account the metallicity evolution of the Universe, we find that the peak of the X-ray luminosity from HMXBs per unit volume, occurs $\sim 0.5\,\rm Gyr$ before the peak of the SFR density. Although the X-ray luminosity coming from HMXBs, per unit SFR, changes by half an order of magnitude over the evolution of the Universe, the observed signature of this effect is masked by the contamination of LMXBs, making the X-ray luminosity from the whole XRB population per unit SFR approximately constant with redshift.

\item Finally, we found that at $z \lesssim 2.5$, where the emission from LMXBs dominates the total X-ray luminosity, the evolution of the X-ray luminosity per unit stellar mass is related to the average age of the LMXB population, while at  $z \gtrsim 2.5$, the evolution of X-ray luminosity per unit stellar mass can be a probe of the evolution of the specific SFR with redshift.

\end{enumerate}

This first PS study of the evolution of XRB populations across comic time highlights the importance of the study of X-ray luminous normal galaxies at high redshift. Furthermore, it lays the ground for future work that will study role played by XRBs in the formation and evolution of galaxies through feedback processes \citep[e.g.][]{Mirabel2011,Zoltan2011,JS2012}.

\acknowledgements 

The authors thank the anonymous referee whose careful report has helped to improve this paper. TF acknowledges support from the CfA and the ITC prize fellowship programs. This work was partially supported from NASA ADAP grant 09-ADP09-0071 (P.I. AH). BDL thanks the Einstein Fellowship Program. PT acknowledges support through a NASA Postdoctoral Program Fellowship at Goddard Space Flight Center, administered by Oak Ridge Associated Universities through a contract with NASA. KB acknowledges support from MSHE grant N203 404939. Computational resources supporting this work were provided by Northwestern University Quest HPC cluster and the NASA High-End Computing (HEC) Program through the NASA Center for Climate Simulation (NCCS) at Goddard Space Flight Center.

 \newpage
 \appendix
 
 \section{To appear as supplemental online-only material}
 \LongTables
 \begin{deluxetable}{ccccccccc} 
\centering
\tablecolumns{9}
\tabletypesize{\scriptsize}
\tablewidth{0pt}
\tablecaption{A complete list of all PS models used in this work. 
\label{modellist_full}} 
\tablehead{ \colhead{Model} & 
     \colhead{$\alpha_{CE}$\tablenotemark{a}} & 
     \colhead{IMF exponent\tablenotemark{b}} & 
     \colhead{$\eta_{wind}$\tablenotemark{c}} & 
     \colhead{CE-HG\tablenotemark{d}} & 
     \colhead{q distribution\tablenotemark{e}} & 
     \colhead{$\kappa_{DC BH}$\tablenotemark{f}} &
     \colhead{Rank\tablenotemark{g}} &
     \colhead{$\log$(Likelihood ratio)\tablenotemark{h}} 
     }
\startdata
       1  &  0.1  &  -2.35  &  1.0  &  No  &  flat  &  0.0  &        81  &        -6.8439794  \\  
       2  &  0.2  &  -2.35  &  1.0  &  No  &  flat  &  0.0  &       131  &        -13.449294  \\  
       3  &  0.3  &  -2.35  &  1.0  &  No  &  flat  &  0.0  &       149  &        -17.435949  \\  
       4  &  0.5  &  -2.35  &  1.0  &  No  &  flat  &  0.0  &       167  &        -22.784369  \\  
       5  &  0.1  &  -2.7  &  1.0  &  No  &  flat  &  0.0  &        23  &        -1.7703540  \\  
       6  &  0.2  &  -2.7  &  1.0  &  No  &  flat  &  0.0  &        65  &        -5.9952885  \\  
       7  &  0.3  &  -2.7  &  1.0  &  No  &  flat  &  0.0  &        99  &        -8.6721399  \\  
       8  &  0.5  &  -2.7  &  1.0  &  No  &  flat  &  0.0  &       121  &        -11.957086  \\  
       9  &  0.1  &  -2.35  &  2.0  &  No  &  flat  &  0.0  &        29  &        -2.3030907  \\  
      10  &  0.2  &  -2.35  &  2.0  &  No  &  flat  &  0.0  &        93  &        -8.1440863  \\  
      11  &  0.3  &  -2.35  &  2.0  &  No  &  flat  &  0.0  &       113  &        -11.281685  \\  
      12  &  0.5  &  -2.35  &  2.0  &  No  &  flat  &  0.0  &       115  &        -11.393684  \\  
      13  &  0.1  &  -2.7  &  2.0  &  No  &  flat  &  0.0  &         9  &       -0.50583441  \\  
      14  &  0.2  &  -2.7  &  2.0  &  No  &  flat  &  0.0  &        51  &        -4.3697639  \\  
      15  &  0.3  &  -2.7  &  2.0  &  No  &  flat  &  0.0  &        70  &        -6.2810696  \\  
      16  &  0.5  &  -2.7  &  2.0  &  No  &  flat  &  0.0  &        64  &        -5.8452086  \\  
      17  &  0.1  &  -2.35  &  0.25  &  No  &  flat  &  0.0  &       160  &        -20.757946  \\  
      18  &  0.2  &  -2.35  &  0.25  &  No  &  flat  &  0.0  &       175  &        -27.096760  \\  
      19  &  0.3  &  -2.35  &  0.25  &  No  &  flat  &  0.0  &       183  &        -32.066400  \\  
      20  &  0.5  &  -2.35  &  0.25  &  No  &  flat  &  0.0  &       190  &        -40.521407  \\  
      21  &  0.1  &  -2.7  &  0.25  &  No  &  flat  &  0.0  &       101  &        -8.9727405  \\  
      22  &  0.2  &  -2.7  &  0.25  &  No  &  flat  &  0.0  &       137  &        -14.797751  \\  
      23  &  0.3  &  -2.7  &  0.25  &  No  &  flat  &  0.0  &       151  &        -18.195038  \\  
      24  &  0.5  &  -2.7  &  0.25  &  No  &  flat  &  0.0  &       168  &        -23.306565  \\  
      25  &  0.1  &  -2.35  &  1.0  &  Yes  &  flat  &  0.0  &        94  &        -8.1863846  \\  
      26  &  0.2  &  -2.35  &  1.0  &  Yes  &  flat  &  0.0  &       138  &        -14.962405  \\  
      27  &  0.3  &  -2.35  &  1.0  &  Yes  &  flat  &  0.0  &       156  &        -19.167849  \\  
      28  &  0.5  &  -2.35  &  1.0  &  Yes  &  flat  &  0.0  &       174  &        -26.883317  \\  
      29  &  0.1  &  -2.7  &  1.0  &  Yes  &  flat  &  0.0  &        24  &        -1.8904716  \\  
      30  &  0.2  &  -2.7  &  1.0  &  Yes  &  flat  &  0.0  &        82  &        -6.8828256  \\  
      31  &  0.3  &  -2.7  &  1.0  &  Yes  &  flat  &  0.0  &       108  &        -10.076327  \\  
      32  &  0.5  &  -2.7  &  1.0  &  Yes  &  flat  &  0.0  &       134  &        -14.094516  \\  
      33  &  0.1  &  -2.35  &  2.0  &  Yes  &  flat  &  0.0  &        30  &        -2.3061373  \\  
      34  &  0.2  &  -2.35  &  2.0  &  Yes  &  flat  &  0.0  &       100  &        -8.8750378  \\  
      35  &  0.3  &  -2.35  &  2.0  &  Yes  &  flat  &  0.0  &       122  &        -12.034528  \\  
      36  &  0.5  &  -2.35  &  2.0  &  Yes  &  flat  &  0.0  &       128  &        -13.072784  \\  
      37  &  0.1  &  -2.7  &  2.0  &  Yes  &  flat  &  0.0  &         7  &       -0.45466309  \\  
      38  &  0.2  &  -2.7  &  2.0  &  Yes  &  flat  &  0.0  &        49  &        -4.1820059  \\  
      39  &  0.3  &  -2.7  &  2.0  &  Yes  &  flat  &  0.0  &        76  &        -6.4496818  \\  
      40  &  0.5  &  -2.7  &  2.0  &  Yes  &  flat  &  0.0  &        74  &        -6.4069133  \\  
      41  &  0.1  &  -2.35  &  0.25  &  Yes  &  flat  &  0.0  &       159  &        -20.659322  \\  
      42  &  0.2  &  -2.35  &  0.25  &  Yes  &  flat  &  0.0  &       182  &        -30.351627  \\  
      43  &  0.3  &  -2.35  &  0.25  &  Yes  &  flat  &  0.0  &       187  &        -36.893741  \\  
      44  &  0.5  &  -2.35  &  0.25  &  Yes  &  flat  &  0.0  &       192  &        -48.616001  \\  
      45  &  0.1  &  -2.7  &  0.25  &  Yes  &  flat  &  0.0  &       106  &        -9.8900460  \\  
      46  &  0.2  &  -2.7  &  0.25  &  Yes  &  flat  &  0.0  &       143  &        -16.465170  \\  
      47  &  0.3  &  -2.7  &  0.25  &  Yes  &  flat  &  0.0  &       164  &        -21.251989  \\  
      48  &  0.5  &  -2.7  &  0.25  &  Yes  &  flat  &  0.0  &       179  &        -28.630340  \\  
      49  &  0.1  &  -2.35  &  1.0  &  No  &  flat  &  0.1  &        58  &        -5.4944514  \\  
      50  &  0.2  &  -2.35  &  1.0  &  No  &  flat  &  0.1  &       119  &        -11.860356  \\  
      51  &  0.3  &  -2.35  &  1.0  &  No  &  flat  &  0.1  &       140  &        -15.604207  \\  
      52  &  0.5  &  -2.35  &  1.0  &  No  &  flat  &  0.1  &       165  &        -21.300323  \\  
      53  &  0.1  &  -2.7  &  1.0  &  No  &  flat  &  0.1  &        20  &        -1.3945094  \\  
      54  &  0.2  &  -2.7  &  1.0  &  No  &  flat  &  0.1  &        59  &        -5.5543889  \\  
      55  &  0.3  &  -2.7  &  1.0  &  No  &  flat  &  0.1  &        96  &        -8.3582206  \\  
      56  &  0.5  &  -2.7  &  1.0  &  No  &  flat  &  0.1  &       109  &        -10.902530  \\  
      57  &  0.1  &  -2.35  &  2.0  &  No  &  flat  &  0.1  &        28  &        -2.2146716  \\  
      58  &  0.2  &  -2.35  &  2.0  &  No  &  flat  &  0.1  &        92  &        -8.0877140  \\  
      59  &  0.3  &  -2.35  &  2.0  &  No  &  flat  &  0.1  &       112  &        -11.149264  \\  
      60  &  0.5  &  -2.35  &  2.0  &  No  &  flat  &  0.1  &       111  &        -11.047687  \\  
      61  &  0.1  &  -2.7  &  2.0  &  No  &  flat  &  0.1  &        16  &        -1.1282696  \\  
      62  &  0.2  &  -2.7  &  2.0  &  No  &  flat  &  0.1  &        52  &        -4.4089974  \\  
      63  &  0.3  &  -2.7  &  2.0  &  No  &  flat  &  0.1  &        77  &        -6.5627561  \\  
      64  &  0.5  &  -2.7  &  2.0  &  No  &  flat  &  0.1  &        72  &        -6.3313205  \\  
      65  &  0.1  &  -2.35  &  0.25  &  No  &  flat  &  0.1  &       147  &        -16.970832  \\  
      66  &  0.2  &  -2.35  &  0.25  &  No  &  flat  &  0.1  &       170  &        -24.339411  \\  
      67  &  0.3  &  -2.35  &  0.25  &  No  &  flat  &  0.1  &       181  &        -28.872144  \\  
      68  &  0.5  &  -2.35  &  0.25  &  No  &  flat  &  0.1  &       188  &        -37.108159  \\  
      69  &  0.1  &  -2.7  &  0.25  &  No  &  flat  &  0.1  &        79  &        -6.6803483  \\  
      70  &  0.2  &  -2.7  &  0.25  &  No  &  flat  &  0.1  &       123  &        -12.306121  \\  
      71  &  0.3  &  -2.7  &  0.25  &  No  &  flat  &  0.1  &       141  &        -15.707806  \\  
      72  &  0.5  &  -2.7  &  0.25  &  No  &  flat  &  0.1  &       163  &        -21.235448  \\  
      73  &  0.1  &  -2.35  &  1.0  &  Yes  &  flat  &  0.1  &        71  &        -6.3013718  \\  
      74  &  0.2  &  -2.35  &  1.0  &  Yes  &  flat  &  0.1  &       129  &        -13.103548  \\  
      75  &  0.3  &  -2.35  &  1.0  &  Yes  &  flat  &  0.1  &       153  &        -18.401280  \\  
      76  &  0.5  &  -2.35  &  1.0  &  Yes  &  flat  &  0.1  &       172  &        -25.574837  \\  
      77  &  0.1  &  -2.7  &  1.0  &  Yes  &  flat  &  0.1  &        21  &        -1.5355082  \\  
      78  &  0.2  &  -2.7  &  1.0  &  Yes  &  flat  &  0.1  &        73  &        -6.4029360  \\  
      79  &  0.3  &  -2.7  &  1.0  &  Yes  &  flat  &  0.1  &       103  &        -9.3432712  \\  
      80  &  0.5  &  -2.7  &  1.0  &  Yes  &  flat  &  0.1  &       130  &        -13.387239  \\  
      81  &  0.1  &  -2.35  &  2.0  &  Yes  &  flat  &  0.1  &        26  &        -2.1743665  \\  
      82  &  0.2  &  -2.35  &  2.0  &  Yes  &  flat  &  0.1  &        98  &        -8.5541878  \\  
      83  &  0.3  &  -2.35  &  2.0  &  Yes  &  flat  &  0.1  &       118  &        -11.762885  \\  
      84  &  0.5  &  -2.35  &  2.0  &  Yes  &  flat  &  0.1  &       127  &        -13.021688  \\  
      85  &  0.1  &  -2.7  &  2.0  &  Yes  &  flat  &  0.1  &        12  &       -0.78509022  \\  
      86  &  0.2  &  -2.7  &  2.0  &  Yes  &  flat  &  0.1  &        50  &        -4.2847568  \\  
      87  &  0.3  &  -2.7  &  2.0  &  Yes  &  flat  &  0.1  &        68  &        -6.2687203  \\  
      88  &  0.5  &  -2.7  &  2.0  &  Yes  &  flat  &  0.1  &        75  &        -6.4135880  \\  
      89  &  0.1  &  -2.35  &  0.25  &  Yes  &  flat  &  0.1  &       152  &        -18.205247  \\  
      90  &  0.2  &  -2.35  &  0.25  &  Yes  &  flat  &  0.1  &       176  &        -27.999553  \\  
      91  &  0.3  &  -2.35  &  0.25  &  Yes  &  flat  &  0.1  &       186  &        -34.997344  \\  
      92  &  0.5  &  -2.35  &  0.25  &  Yes  &  flat  &  0.1  &       191  &        -46.958394  \\  
      93  &  0.1  &  -2.7  &  0.25  &  Yes  &  flat  &  0.1  &        86  &        -7.1404740  \\  
      94  &  0.2  &  -2.7  &  0.25  &  Yes  &  flat  &  0.1  &       136  &        -14.372125  \\  
      95  &  0.3  &  -2.7  &  0.25  &  Yes  &  flat  &  0.1  &       158  &        -20.101154  \\  
      96  &  0.5  &  -2.7  &  0.25  &  Yes  &  flat  &  0.1  &       178  &        -28.160352  \\  
      97  &  0.1  &  -2.35  &  1.0  &  No  &  twin  &  0.0  &       260  &        -156.87865  \\  
      98  &  0.2  &  -2.35  &  1.0  &  No  &  twin  &  0.0  &       226  &        -118.90171  \\  
      99  &  0.3  &  -2.35  &  1.0  &  No  &  twin  &  0.0  &       227  &        -120.00800  \\  
     100  &  0.5  &  -2.35  &  1.0  &  No  &  twin  &  0.0  &       273  &        -173.98895  \\  
     101  &  0.1  &  -2.7  &  1.0  &  No  &  twin  &  0.0  &       274  &        -177.22909  \\  
     102  &  0.2  &  -2.7  &  1.0  &  No  &  twin  &  0.0  &       234  &        -126.11918  \\  
     103  &  0.3  &  -2.7  &  1.0  &  No  &  twin  &  0.0  &       239  &        -131.17535  \\  
     104  &  0.5  &  -2.7  &  1.0  &  No  &  twin  &  0.0  &       280  &        -189.56799  \\  
     105  &  0.1  &  -2.35  &  2.0  &  No  &  twin  &  0.0  &       276  &        -181.10990  \\  
     106  &  0.2  &  -2.35  &  2.0  &  No  &  twin  &  0.0  &       246  &        -139.37974  \\  
     107  &  0.3  &  -2.35  &  2.0  &  No  &  twin  &  0.0  &       232  &        -124.52035  \\  
     108  &  0.5  &  -2.35  &  2.0  &  No  &  twin  &  0.0  &       279  &        -183.77797  \\  
     109  &  0.1  &  -2.7  &  2.0  &  No  &  twin  &  0.0  &       283  &        -212.54419  \\  
     110  &  0.2  &  -2.7  &  2.0  &  No  &  twin  &  0.0  &       249  &        -141.81642  \\  
     111  &  0.3  &  -2.7  &  2.0  &  No  &  twin  &  0.0  &       252  &        -144.96094  \\  
     112  &  0.5  &  -2.7  &  2.0  &  No  &  twin  &  0.0  &       275  &        -180.61817  \\  
     113  &  0.1  &  -2.35  &  0.25  &  No  &  twin  &  0.0  &       281  &        -190.13124  \\  
     114  &  0.2  &  -2.35  &  0.25  &  No  &  twin  &  0.0  &       209  &        -105.58207  \\  
     115  &  0.3  &  -2.35  &  0.25  &  No  &  twin  &  0.0  &       220  &        -113.36827  \\  
     116  &  0.5  &  -2.35  &  0.25  &  No  &  twin  &  0.0  &       243  &        -136.34902  \\  
     117  &  0.1  &  -2.7  &  0.25  &  No  &  twin  &  0.0  &       271  &        -173.34531  \\  
     118  &  0.2  &  -2.7  &  0.25  &  No  &  twin  &  0.0  &       212  &        -107.65630  \\  
     119  &  0.3  &  -2.7  &  0.25  &  No  &  twin  &  0.0  &       223  &        -115.87870  \\  
     120  &  0.5  &  -2.7  &  0.25  &  No  &  twin  &  0.0  &       259  &        -156.69007  \\  
     121  &  0.1  &  -2.35  &  1.0  &  Yes  &  twin  &  0.0  &       265  &        -166.58676  \\  
     122  &  0.2  &  -2.35  &  1.0  &  Yes  &  twin  &  0.0  &       236  &        -128.34509  \\  
     123  &  0.3  &  -2.35  &  1.0  &  Yes  &  twin  &  0.0  &       202  &        -102.01665  \\  
     124  &  0.5  &  -2.35  &  1.0  &  Yes  &  twin  &  0.0  &       199  &        -99.689704  \\  
     125  &  0.1  &  -2.7  &  1.0  &  Yes  &  twin  &  0.0  &       270  &        -171.79474  \\  
     126  &  0.2  &  -2.7  &  1.0  &  Yes  &  twin  &  0.0  &       245  &        -138.27339  \\  
     127  &  0.3  &  -2.7  &  1.0  &  Yes  &  twin  &  0.0  &       231  &        -123.30591  \\  
     128  &  0.5  &  -2.7  &  1.0  &  Yes  &  twin  &  0.0  &       219  &        -111.26572  \\  
     129  &  0.1  &  -2.35  &  2.0  &  Yes  &  twin  &  0.0  &       269  &        -170.91556  \\  
     130  &  0.2  &  -2.35  &  2.0  &  Yes  &  twin  &  0.0  &       244  &        -138.08749  \\  
     131  &  0.3  &  -2.35  &  2.0  &  Yes  &  twin  &  0.0  &       233  &        -125.38742  \\  
     132  &  0.5  &  -2.35  &  2.0  &  Yes  &  twin  &  0.0  &       204  &        -103.05276  \\  
     133  &  0.1  &  -2.7  &  2.0  &  Yes  &  twin  &  0.0  &       287  &        -238.92812  \\  
     134  &  0.2  &  -2.7  &  2.0  &  Yes  &  twin  &  0.0  &       253  &        -145.13205  \\  
     135  &  0.3  &  -2.7  &  2.0  &  Yes  &  twin  &  0.0  &       250  &        -142.40690  \\  
     136  &  0.5  &  -2.7  &  2.0  &  Yes  &  twin  &  0.0  &       225  &        -117.78790  \\  
     137  &  0.1  &  -2.35  &  0.25  &  Yes  &  twin  &  0.0  &       257  &        -150.66356  \\  
     138  &  0.2  &  -2.35  &  0.25  &  Yes  &  twin  &  0.0  &       206  &        -103.42077  \\  
     139  &  0.3  &  -2.35  &  0.25  &  Yes  &  twin  &  0.0  &       197  &        -99.321415  \\  
     140  &  0.5  &  -2.35  &  0.25  &  Yes  &  twin  &  0.0  &       198  &        -99.520795  \\  
     141  &  0.1  &  -2.7  &  0.25  &  Yes  &  twin  &  0.0  &       268  &        -170.03313  \\  
     142  &  0.2  &  -2.7  &  0.25  &  Yes  &  twin  &  0.0  &       221  &        -115.28205  \\  
     143  &  0.3  &  -2.7  &  0.25  &  Yes  &  twin  &  0.0  &       196  &        -99.065831  \\  
     144  &  0.5  &  -2.7  &  0.25  &  Yes  &  twin  &  0.0  &       203  &        -103.00718  \\  
     145  &  0.1  &  -2.35  &  1.0  &  No  &  twin  &  0.1  &       282  &        -212.05569  \\  
     146  &  0.2  &  -2.35  &  1.0  &  No  &  twin  &  0.1  &       211  &        -107.37148  \\  
     147  &  0.3  &  -2.35  &  1.0  &  No  &  twin  &  0.1  &       215  &        -109.36328  \\  
     148  &  0.5  &  -2.35  &  1.0  &  No  &  twin  &  0.1  &       247  &        -140.54672  \\  
     149  &  0.1  &  -2.7  &  1.0  &  No  &  twin  &  0.1  &       261  &        -158.65389  \\  
     150  &  0.2  &  -2.7  &  1.0  &  No  &  twin  &  0.1  &       240  &        -131.79837  \\  
     151  &  0.3  &  -2.7  &  1.0  &  No  &  twin  &  0.1  &       228  &        -120.63145  \\  
     152  &  0.5  &  -2.7  &  1.0  &  No  &  twin  &  0.1  &       263  &        -163.96495  \\  
     153  &  0.1  &  -2.35  &  2.0  &  No  &  twin  &  0.1  &       285  &        -226.73511  \\  
     154  &  0.2  &  -2.35  &  2.0  &  No  &  twin  &  0.1  &       255  &        -147.27372  \\  
     155  &  0.3  &  -2.35  &  2.0  &  No  &  twin  &  0.1  &       241  &        -131.82939  \\  
     156  &  0.5  &  -2.35  &  2.0  &  No  &  twin  &  0.1  &       278  &        -183.64930  \\  
     157  &  0.1  &  -2.7  &  2.0  &  No  &  twin  &  0.1  &       286  &        -229.31936  \\  
     158  &  0.2  &  -2.7  &  2.0  &  No  &  twin  &  0.1  &       272  &        -173.45129  \\  
     159  &  0.3  &  -2.7  &  2.0  &  No  &  twin  &  0.1  &       256  &        -149.25108  \\  
     160  &  0.5  &  -2.7  &  2.0  &  No  &  twin  &  0.1  &       284  &        -213.83915  \\  
     161  &  0.1  &  -2.35  &  0.25  &  No  &  twin  &  0.1  &       267  &        -169.73810  \\  
     162  &  0.2  &  -2.35  &  0.25  &  No  &  twin  &  0.1  &       200  &        -101.25795  \\  
     163  &  0.3  &  -2.35  &  0.25  &  No  &  twin  &  0.1  &       216  &        -109.61989  \\  
     164  &  0.5  &  -2.35  &  0.25  &  No  &  twin  &  0.1  &       237  &        -128.62607  \\  
     165  &  0.1  &  -2.7  &  0.25  &  No  &  twin  &  0.1  &       254  &        -146.26185  \\  
     166  &  0.2  &  -2.7  &  0.25  &  No  &  twin  &  0.1  &       210  &        -105.95448  \\  
     167  &  0.3  &  -2.7  &  0.25  &  No  &  twin  &  0.1  &       235  &        -127.40935  \\  
     168  &  0.5  &  -2.7  &  0.25  &  No  &  twin  &  0.1  &       251  &        -142.65248  \\  
     169  &  0.1  &  -2.35  &  1.0  &  Yes  &  twin  &  0.1  &       277  &        -182.07044  \\  
     170  &  0.2  &  -2.35  &  1.0  &  Yes  &  twin  &  0.1  &       224  &        -116.16233  \\  
     171  &  0.3  &  -2.35  &  1.0  &  Yes  &  twin  &  0.1  &       217  &        -110.39158  \\  
     172  &  0.5  &  -2.35  &  1.0  &  Yes  &  twin  &  0.1  &       201  &        -101.26795  \\  
     173  &  0.1  &  -2.7  &  1.0  &  Yes  &  twin  &  0.1  &       213  &        -108.61106  \\  
     174  &  0.2  &  -2.7  &  1.0  &  Yes  &  twin  &  0.1  &       230  &        -122.80642  \\  
     175  &  0.3  &  -2.7  &  1.0  &  Yes  &  twin  &  0.1  &       222  &        -115.68776  \\  
     176  &  0.5  &  -2.7  &  1.0  &  Yes  &  twin  &  0.1  &       218  &        -111.09642  \\  
     177  &  0.1  &  -2.35  &  2.0  &  Yes  &  twin  &  0.1  &       262  &        -158.68953  \\  
     178  &  0.2  &  -2.35  &  2.0  &  Yes  &  twin  &  0.1  &       248  &        -141.13179  \\  
     179  &  0.3  &  -2.35  &  2.0  &  Yes  &  twin  &  0.1  &       238  &        -130.61604  \\  
     180  &  0.5  &  -2.35  &  2.0  &  Yes  &  twin  &  0.1  &       207  &        -103.52411  \\  
     181  &  0.1  &  -2.7  &  2.0  &  Yes  &  twin  &  0.1  &       288  &        -254.52814  \\  
     182  &  0.2  &  -2.7  &  2.0  &  Yes  &  twin  &  0.1  &       264  &        -166.30555  \\  
     183  &  0.3  &  -2.7  &  2.0  &  Yes  &  twin  &  0.1  &       242  &        -132.91527  \\  
     184  &  0.5  &  -2.7  &  2.0  &  Yes  &  twin  &  0.1  &       229  &        -121.68847  \\  
     185  &  0.1  &  -2.35  &  0.25  &  Yes  &  twin  &  0.1  &       266  &        -169.59003  \\  
     186  &  0.2  &  -2.35  &  0.25  &  Yes  &  twin  &  0.1  &       208  &        -103.86318  \\  
     187  &  0.3  &  -2.35  &  0.25  &  Yes  &  twin  &  0.1  &       194  &        -94.819491  \\  
     188  &  0.5  &  -2.35  &  0.25  &  Yes  &  twin  &  0.1  &       193  &        -92.661723  \\  
     189  &  0.1  &  -2.7  &  0.25  &  Yes  &  twin  &  0.1  &       258  &        -156.51391  \\  
     190  &  0.2  &  -2.7  &  0.25  &  Yes  &  twin  &  0.1  &       214  &        -108.86360  \\  
     191  &  0.3  &  -2.7  &  0.25  &  Yes  &  twin  &  0.1  &       205  &        -103.13924  \\  
     192  &  0.5  &  -2.7  &  0.25  &  Yes  &  twin  &  0.1  &       195  &        -98.010507  \\  
     193  &  0.1  &  -2.35  &  1.0  &  No  &  50-50  &  0.0  &        56  &        -4.6196089  \\  
     194  &  0.2  &  -2.35  &  1.0  &  No  &  50-50  &  0.0  &       102  &        -8.9811857  \\  
     195  &  0.3  &  -2.35  &  1.0  &  No  &  50-50  &  0.0  &       120  &        -11.949863  \\  
     196  &  0.5  &  -2.35  &  1.0  &  No  &  50-50  &  0.0  &       142  &        -16.147741  \\  
     197  &  0.1  &  -2.7  &  1.0  &  No  &  50-50  &  0.0  &        13  &       -0.80759106  \\  
     198  &  0.2  &  -2.7  &  1.0  &  No  &  50-50  &  0.0  &        39  &        -2.8435160  \\  
     199  &  0.3  &  -2.7  &  1.0  &  No  &  50-50  &  0.0  &        55  &        -4.5791773  \\  
     200  &  0.5  &  -2.7  &  1.0  &  No  &  50-50  &  0.0  &        83  &        -6.9038795  \\  
     201  &  0.1  &  -2.35  &  2.0  &  No  &  50-50  &  0.0  &        10  &       -0.59598324  \\  
     202  &  0.2  &  -2.35  &  2.0  &  No  &  50-50  &  0.0  &        48  &        -3.9985781  \\  
     203  &  0.3  &  -2.35  &  2.0  &  No  &  50-50  &  0.0  &        66  &        -6.1997086  \\  
     204  &  0.5  &  -2.35  &  2.0  &  No  &  50-50  &  0.0  &        69  &        -6.2738874  \\  
     205  &  0.1  &  -2.7  &  2.0  &  No  &  50-50  &  0.0  &         4  &       -0.13235138  \\  
     206  &  0.2  &  -2.7  &  2.0  &  No  &  50-50  &  0.0  &        18  &        -1.3037351  \\  
     207  &  0.3  &  -2.7  &  2.0  &  No  &  50-50  &  0.0  &        32  &        -2.4529087  \\  
     208  &  0.5  &  -2.7  &  2.0  &  No  &  50-50  &  0.0  &        27  &        -2.1765207  \\  
     209  &  0.1  &  -2.35  &  0.25  &  No  &  50-50  &  0.0  &       145  &        -16.868460  \\  
     210  &  0.2  &  -2.35  &  0.25  &  No  &  50-50  &  0.0  &       166  &        -21.835832  \\  
     211  &  0.3  &  -2.35  &  0.25  &  No  &  50-50  &  0.0  &       173  &        -25.882512  \\  
     212  &  0.5  &  -2.35  &  0.25  &  No  &  50-50  &  0.0  &       184  &        -32.902336  \\  
     213  &  0.1  &  -2.7  &  0.25  &  No  &  50-50  &  0.0  &        84  &        -6.9688649  \\  
     214  &  0.2  &  -2.7  &  0.25  &  No  &  50-50  &  0.0  &       110  &        -10.931015  \\  
     215  &  0.3  &  -2.7  &  0.25  &  No  &  50-50  &  0.0  &       132  &        -13.487211  \\  
     216  &  0.5  &  -2.7  &  0.25  &  No  &  50-50  &  0.0  &       150  &        -17.481734  \\  
     217  &  0.1  &  -2.35  &  1.0  &  Yes  &  50-50  &  0.0  &        60  &        -5.5770783  \\  
     218  &  0.2  &  -2.35  &  1.0  &  Yes  &  50-50  &  0.0  &       107  &        -9.9479274  \\  
     219  &  0.3  &  -2.35  &  1.0  &  Yes  &  50-50  &  0.0  &       126  &        -12.981021  \\  
     220  &  0.5  &  -2.35  &  1.0  &  Yes  &  50-50  &  0.0  &       154  &        -18.960596  \\  
     221  &  0.1  &  -2.7  &  1.0  &  Yes  &  50-50  &  0.0  &        15  &        -1.0050246  \\  
     222  &  0.2  &  -2.7  &  1.0  &  Yes  &  50-50  &  0.0  &        42  &        -3.4137746  \\  
     223  &  0.3  &  -2.7  &  1.0  &  Yes  &  50-50  &  0.0  &        57  &        -5.4813514  \\  
     224  &  0.5  &  -2.7  &  1.0  &  Yes  &  50-50  &  0.0  &        95  &        -8.3259810  \\  
     225  &  0.1  &  -2.35  &  2.0  &  Yes  &  50-50  &  0.0  &        11  &       -0.75550478  \\  
     226  &  0.2  &  -2.35  &  2.0  &  Yes  &  50-50  &  0.0  &        54  &        -4.5531285  \\  
     227  &  0.3  &  -2.35  &  2.0  &  Yes  &  50-50  &  0.0  &        80  &        -6.7609447  \\  
     228  &  0.5  &  -2.35  &  2.0  &  Yes  &  50-50  &  0.0  &        89  &        -7.4758766  \\  
     229  &  0.1  &  -2.7  &  2.0  &  Yes  &  50-50  &  0.0  &         2  &      -0.035048515  \\  
     230  &  0.2  &  -2.7  &  2.0  &  Yes  &  50-50  &  0.0  &        17  &        -1.2284989  \\  
     231  &  0.3  &  -2.7  &  2.0  &  Yes  &  50-50  &  0.0  &        37  &        -2.6011593  \\  
     232  &  0.5  &  -2.7  &  2.0  &  Yes  &  50-50  &  0.0  &        35  &        -2.5733454  \\  
     233  &  0.1  &  -2.35  &  0.25  &  Yes  &  50-50  &  0.0  &       144  &        -16.735734  \\  
     234  &  0.2  &  -2.35  &  0.25  &  Yes  &  50-50  &  0.0  &       169  &        -23.741728  \\  
     235  &  0.3  &  -2.35  &  0.25  &  Yes  &  50-50  &  0.0  &       180  &        -28.856645  \\  
     236  &  0.5  &  -2.35  &  0.25  &  Yes  &  50-50  &  0.0  &       189  &        -38.542077  \\  
     237  &  0.1  &  -2.7  &  0.25  &  Yes  &  50-50  &  0.0  &        91  &        -7.5587255  \\  
     238  &  0.2  &  -2.7  &  0.25  &  Yes  &  50-50  &  0.0  &       117  &        -11.759989  \\  
     239  &  0.3  &  -2.7  &  0.25  &  Yes  &  50-50  &  0.0  &       139  &        -15.228415  \\  
     240  &  0.5  &  -2.7  &  0.25  &  Yes  &  50-50  &  0.0  &       161  &        -20.960307  \\  
     241  &  0.1  &  -2.35  &  1.0  &  No  &  50-50  &  0.1  &        31  &        -2.3582808  \\  
     242  &  0.2  &  -2.35  &  1.0  &  No  &  50-50  &  0.1  &        78  &        -6.6277666  \\  
     243  &  0.3  &  -2.35  &  1.0  &  No  &  50-50  &  0.1  &       104  &        -9.4199752  \\  
     244  &  0.5  &  -2.35  &  1.0  &  No  &  50-50  &  0.1  &       133  &        -13.899182  \\  
     245  &  0.1  &  -2.7  &  1.0  &  No  &  50-50  &  0.1  &         1  &         0.0000000  \\  
     246  &  0.2  &  -2.7  &  1.0  &  No  &  50-50  &  0.1  &        25  &        -2.0332112  \\  
     247  &  0.3  &  -2.7  &  1.0  &  No  &  50-50  &  0.1  &        45  &        -3.8604629  \\  
     248  &  0.5  &  -2.7  &  1.0  &  No  &  50-50  &  0.1  &        61  &        -5.6476472  \\  
     249  &  0.1  &  -2.35  &  2.0  &  No  &  50-50  &  0.1  &         5  &       -0.28666705  \\  
     250  &  0.2  &  -2.35  &  2.0  &  No  &  50-50  &  0.1  &        44  &        -3.6659021  \\  
     251  &  0.3  &  -2.35  &  2.0  &  No  &  50-50  &  0.1  &        63  &        -5.8169795  \\  
     252  &  0.5  &  -2.35  &  2.0  &  No  &  50-50  &  0.1  &        62  &        -5.7460627  \\  
     253  &  0.1  &  -2.7  &  2.0  &  No  &  50-50  &  0.1  &        14  &       -0.90617906  \\  
     254  &  0.2  &  -2.7  &  2.0  &  No  &  50-50  &  0.1  &        22  &        -1.6301615  \\  
     255  &  0.3  &  -2.7  &  2.0  &  No  &  50-50  &  0.1  &        40  &        -2.8798898  \\  
     256  &  0.5  &  -2.7  &  2.0  &  No  &  50-50  &  0.1  &        38  &        -2.7683023  \\  
     257  &  0.1  &  -2.35  &  0.25  &  No  &  50-50  &  0.1  &       116  &        -11.499357  \\  
     258  &  0.2  &  -2.35  &  0.25  &  No  &  50-50  &  0.1  &       148  &        -17.358444  \\  
     259  &  0.3  &  -2.35  &  0.25  &  No  &  50-50  &  0.1  &       162  &        -21.119200  \\  
     260  &  0.5  &  -2.35  &  0.25  &  No  &  50-50  &  0.1  &       177  &        -28.019414  \\  
     261  &  0.1  &  -2.7  &  0.25  &  No  &  50-50  &  0.1  &        43  &        -3.4548391  \\  
     262  &  0.2  &  -2.7  &  0.25  &  No  &  50-50  &  0.1  &        87  &        -7.2923632  \\  
     263  &  0.3  &  -2.7  &  0.25  &  No  &  50-50  &  0.1  &       105  &        -9.8528845  \\  
     264  &  0.5  &  -2.7  &  0.25  &  No  &  50-50  &  0.1  &       135  &        -14.206096  \\  
     265  &  0.1  &  -2.35  &  1.0  &  Yes  &  50-50  &  0.1  &        41  &        -3.1652049  \\  
     266  &  0.2  &  -2.35  &  1.0  &  Yes  &  50-50  &  0.1  &        90  &        -7.5318220  \\  
     267  &  0.3  &  -2.35  &  1.0  &  Yes  &  50-50  &  0.1  &       114  &        -11.360538  \\  
     268  &  0.5  &  -2.35  &  1.0  &  Yes  &  50-50  &  0.1  &       146  &        -16.893279  \\  
     269  &  0.1  &  -2.7  &  1.0  &  Yes  &  50-50  &  0.1  &         3  &       -0.11226917  \\  
     270  &  0.2  &  -2.7  &  1.0  &  Yes  &  50-50  &  0.1  &        34  &        -2.5620627  \\  
     271  &  0.3  &  -2.7  &  1.0  &  Yes  &  50-50  &  0.1  &        53  &        -4.4550489  \\  
     272  &  0.5  &  -2.7  &  1.0  &  Yes  &  50-50  &  0.1  &        88  &        -7.3062420  \\  
     273  &  0.1  &  -2.35  &  2.0  &  Yes  &  50-50  &  0.1  &         6  &       -0.28926393  \\  
     274  &  0.2  &  -2.35  &  2.0  &  Yes  &  50-50  &  0.1  &        47  &        -3.9974158  \\  
     275  &  0.3  &  -2.35  &  2.0  &  Yes  &  50-50  &  0.1  &        67  &        -6.2402162  \\  
     276  &  0.5  &  -2.35  &  2.0  &  Yes  &  50-50  &  0.1  &        85  &        -7.1175851  \\  
     277  &  0.1  &  -2.7  &  2.0  &  Yes  &  50-50  &  0.1  &         8  &       -0.46019542  \\  
     278  &  0.2  &  -2.7  &  2.0  &  Yes  &  50-50  &  0.1  &        19  &        -1.3618415  \\  
     279  &  0.3  &  -2.7  &  2.0  &  Yes  &  50-50  &  0.1  &        33  &        -2.5161509  \\  
     280  &  0.5  &  -2.7  &  2.0  &  Yes  &  50-50  &  0.1  &        36  &        -2.5963346  \\  
     281  &  0.1  &  -2.35  &  0.25  &  Yes  &  50-50  &  0.1  &       124  &        -12.491142  \\  
     282  &  0.2  &  -2.35  &  0.25  &  Yes  &  50-50  &  0.1  &       157  &        -19.571515  \\  
     283  &  0.3  &  -2.35  &  0.25  &  Yes  &  50-50  &  0.1  &       171  &        -25.070431  \\  
     284  &  0.5  &  -2.35  &  0.25  &  Yes  &  50-50  &  0.1  &       185  &        -34.930124  \\  
     285  &  0.1  &  -2.7  &  0.25  &  Yes  &  50-50  &  0.1  &        46  &        -3.8622306  \\  
     286  &  0.2  &  -2.7  &  0.25  &  Yes  &  50-50  &  0.1  &        97  &        -8.5070409  \\  
     287  &  0.3  &  -2.7  &  0.25  &  Yes  &  50-50  &  0.1  &       125  &        -12.678477  \\  
     288  &  0.5  &  -2.7  &  0.25  &  Yes  &  50-50  &  0.1  &       155  &        -18.965882                                                               
\enddata

\tablenotetext{a}{CE efficiency parameter}
\tablenotetext{b}{Exponent of the high-mass power law component of the IMF: \citet{Kroupa2001} (-2.35) or \citet{KW2003} (-2.7).}
\tablenotetext{c}{Stellar wind strength parameter with which the ``standard'' \citep{Belczynski2010} stellar wind prescription is multiplied.}
\tablenotetext{d}{Yes: all possible outcomes of a CE event with a HG donor are allowed, No: A CE with a HG donor star will always result to a merger.}
\tablenotetext{e}{Binary mass ratio distribution. ``50-50'' indicates half of the binaries originate from a ``twin binary'' distribution and half from flat mass ratio distribution.} 
\tablenotetext{f}{Parameter with which the ``standard'' \citet{Hobbs2005} kick distribution is multiplied for BHs formed though a SN explosion with negligible ejected mass.}
\tablenotetext{g}{ Ranking of the model in our statistical comparison.}
\tablenotetext{h}{ Ratio of the likelihood of the observations given a model to our maximum likelihood model ($\log{L(O|M_i)}-\log{L(O|M_{ref})}$).}
\end{deluxetable}


\begin{thebibliography}{80}
\expandafter\ifx\csname natexlab\endcsname\relax\def\natexlab#1{#1}\fi

\bibitem[{Abt(1983)}]{Abt1983}
Abt, H.~A. 1983, IN: Annual review of astronomy and astrophysics. Volume 21
  (A84-10851 01-90). Palo Alto, 21, 343

\bibitem[{Basu-Zych {et~al.}(2012)Basu-Zych, Lehmer, Hornschemeier, Bouwens,
  Fragos, Oesch, Belczynski, Brandt, Kalogera, Luo, Miller, Mullaney,
  Tzanavaris, Xue, \& Zezas}]{BasuZych2012}
Basu-Zych, A.~R., Lehmer, B.~D., Hornschemeier, A.~E., Bouwens, R.~J., Fragos,
  T., Oesch, P.~A., Belczynski, K., Brandt, W.~N., Kalogera, V., Luo, B.,
  Miller, N., Mullaney, J.~R., Tzanavaris, P., Xue, Y., \& Zezas, A. 2012,
  arXiv, 1210, 3357

\bibitem[{Bauer {et~al.}(2002)Bauer, Alexander, Brandt, Hornschemeier, Miyaji,
  Garmire, Schneider, Bautz, Chartas, Griffiths, \& Sargent}]{Bauer2002}
Bauer, F.~E., Alexander, D.~M., Brandt, W.~N., Hornschemeier, A.~E., Miyaji,
  T., Garmire, G.~P., Schneider, D.~P., Bautz, M.~W., Chartas, G., Griffiths,
  R.~E., \& Sargent, W.~L.~W. 2002, \aj, 123, 1163

\bibitem[{Belczynski {et~al.}(2010)Belczynski, Bulik, Fryer, Ruiter, Valsecchi,
  Vink, \& Hurley}]{Belczynski2010}
Belczynski, K., Bulik, T., Fryer, C.~L., Ruiter, A., Valsecchi, F., Vink,
  J.~S., \& Hurley, J.~R. 2010, The Astrophysical Journal, 714, 1217

\bibitem[{Belczynski {et~al.}(2002)Belczynski, kalogera, \&
  Bulik}]{Belczynski2002}
Belczynski, K., kalogera, V., \& Bulik, T. 2002, \apj, 572, 407

\bibitem[{Belczynski {et~al.}(2008)Belczynski, Kalogera, Rasio, Taam, Zezas,
  Bulik, Maccarone, \& Ivanova}]{Belczynski2008}
Belczynski, K., Kalogera, V., Rasio, F.~A., Taam, R.~E., Zezas, A., Bulik, T.,
  Maccarone, T.~J., \& Ivanova, N. 2008, \apjs, 174, 223

\bibitem[{Belczynski {et~al.}(2004)Belczynski, kalogera, Zezas, \&
  Fabbiano}]{Belczynski2004}
Belczynski, K., kalogera, V., Zezas, A., \& Fabbiano, G. 2004, \apjl, 601, L147

\bibitem[{Belczynski \& Taam(2004)}]{Belczynski2004b}
Belczynski, K. \& Taam, R.~E. 2004, The Astrophysical Journal, 616, 1159

\bibitem[{Belczynski {et~al.}(2007)Belczynski, Taam, kalogera, Rasio, \&
  Bulik}]{Belczynski2007}
Belczynski, K., Taam, R.~E., kalogera, V., Rasio, F.~A., \& Bulik, T. 2007,
  \apj, 662, 504

\bibitem[{Belczynski {et~al.}(2012)Belczynski, Wiktorowicz, Fryer, Holz, \&
  Kalogera}]{Belczynski2012}
Belczynski, K., Wiktorowicz, G., Fryer, C.~L., Holz, D.~E., \& Kalogera, V.
  2012, The Astrophysical Journal, 757, 91

\bibitem[{Bell {et~al.}(2003)Bell, McIntosh, Katz, \& Weinberg}]{Bell2003}
Bell, E.~F., McIntosh, D.~H., Katz, N., \& Weinberg, M.~D. 2003, \apjs, 149,
  289

\bibitem[{Boroson {et~al.}(2011)Boroson, Kim, \& Fabbiano}]{Boroson2011}
Boroson, B., Kim, D.-W., \& Fabbiano, G. 2011, The Astrophysical Journal, 729,
  12

\bibitem[{Bouwens {et~al.}(2007)Bouwens, Illingworth, Franx, \&
  Ford}]{Bouwens2007}
Bouwens, R.~J., Illingworth, G.~D., Franx, M., \& Ford, H. 2007, The
  Astrophysical Journal, 670, 928

\bibitem[{Boylan-Kolchin {et~al.}(2009)Boylan-Kolchin, Springel, White,
  Jenkins, \& Lemson}]{Boylan2009}
Boylan-Kolchin, M., Springel, V., White, S.~D.~M., Jenkins, A., \& Lemson, G.
  2009, \mnras, 398, 1150

\bibitem[{Brandt \& Hasinger(2005)}]{BH2005}
Brandt, W.~N. \& Hasinger, G. 2005, \araa, 43, 827

\bibitem[{Caballero-Garc{\'\i}a {et~al.}(2009)Caballero-Garc{\'\i}a, Miller,
  Trigo, Kuulkers, Fabian, Mas-Hesse, Steeghs, \& van~der Klis}]{Caballero2009}
Caballero-Garc{\'\i}a, M.~D., Miller, J.~M., Trigo, M.~D., Kuulkers, E.,
  Fabian, A.~C., Mas-Hesse, J.~M., Steeghs, D., \& van~der Klis, M. 2009, The
  Astrophysical Journal, 692, 1339

\bibitem[{Dijkstra {et~al.}(2012)Dijkstra, Gilfanov, Loeb, \&
  Sunyaev}]{Dijkstra2012}
Dijkstra, M., Gilfanov, M., Loeb, A., \& Sunyaev, R. 2012, \mnras, 421, 213

\bibitem[{Dominik {et~al.}(2012)Dominik, Belczynski, Fryer, Holz, Berti, Bulik,
  Mandel, \& O'Shaughnessy}]{Dominik2012}
Dominik, M., Belczynski, K., Fryer, C., Holz, D.~E., Berti, E., Bulik, T.,
  Mandel, I., \& O'Shaughnessy, R. 2012, The Astrophysical Journal, 759, 52

\bibitem[{Fragos {et~al.}(2008)Fragos, kalogera, Belczynski, Fabbiano, Kim,
  Brassington, Angelini, Davies, Gallagher, King, Pellegrini, Trinchieri, Zepf,
  Kundu, \& Zezas}]{Fragos2008}
Fragos, T., kalogera, V., Belczynski, K., Fabbiano, G., Kim, D.-W.,
  Brassington, N.~J., Angelini, L., Davies, R.~L., Gallagher, J.~S., King,
  A.~R., Pellegrini, S., Trinchieri, G., Zepf, S.~E., Kundu, A., \& Zezas, A.
  2008, The Astrophysical Journal, 683, 346

\bibitem[{Fragos {et~al.}(2009)Fragos, kalogera, Willems, Belczynski, Fabbiano,
  Brassington, Kim, Angelini, Davies, Gallagher, King, Pellegrini, Trinchieri,
  Zepf, \& Zezas}]{Fragos2009}
Fragos, T., kalogera, V., Willems, B., Belczynski, K., Fabbiano, G.,
  Brassington, N.~J., Kim, D.-W., Angelini, L., Davies, R.~L., Gallagher,
  J.~S., King, A.~R., Pellegrini, S., Trinchieri, G., Zepf, S.~E., \& Zezas, A.
  2009, \apjl, 702, L143

\bibitem[{Fragos {et~al.}(2010)Fragos, Tremmel, Rantsiou, \&
  Belczynski}]{Fragos2010}
Fragos, T., Tremmel, M., Rantsiou, E., \& Belczynski, K. 2010, \apjl, 719, L79

\bibitem[{Fryer {et~al.}(2012)Fryer, Belczynski, Wiktorowicz, Dominik,
  Kalogera, \& Holz}]{Fryer2012}
Fryer, C.~L., Belczynski, K., Wiktorowicz, G., Dominik, M., Kalogera, V., \&
  Holz, D.~E. 2012, The Astrophysical Journal, 749, 91

\bibitem[{Ghosh \& White(2001)}]{GW2001}
Ghosh, P. \& White, N.~E. 2001, \apjl, 559, L97

\bibitem[{Gilfanov(2004)}]{Gilfanov2004b}
Gilfanov, M. 2004, \mnras, 349, 146

\bibitem[{Gilfanov {et~al.}(2004)Gilfanov, Grimm, \& Sunyaev}]{Gilfanov2004a}
Gilfanov, M., Grimm, H.-J., \& Sunyaev, R. 2004, \mnras, 347, L57

\bibitem[{Gou {et~al.}(2009)Gou, McClintock, Liu, Narayan, Steiner, Remillard,
  Orosz, Davis, Ebisawa, \& Schlegel}]{Gou2009}
Gou, L., McClintock, J.~E., Liu, J., Narayan, R., Steiner, J.~F., Remillard,
  R.~A., Orosz, J.~A., Davis, S.~W., Ebisawa, K., \& Schlegel, E.~M. 2009, The
  Astrophysical Journal, 701, 1076

\bibitem[{Grimm {et~al.}(2002)Grimm, Gilfanov, \& Sunyaev}]{GGS2002}
Grimm, H.-J., Gilfanov, M., \& Sunyaev, R. 2002, \aap, 391, 923

\bibitem[{Guo {et~al.}(2011)Guo, White, Boylan-Kolchin, De~Lucia, Kauffmann,
  Lemson, Li, Springel, \& Weinmann}]{Guo2011}
Guo, Q., White, S., Boylan-Kolchin, M., De~Lucia, G., Kauffmann, G., Lemson,
  G., Li, C., Springel, V., \& Weinmann, S. 2011, \mnras, 413, 101

\bibitem[{Haiman(2011)}]{Zoltan2011}
Haiman, Z. 2011, Nature, 472, 47

\bibitem[{Heggie(1975)}]{Heggie1975}
Heggie, D.~C. 1975, \mnras, 173, 729

\bibitem[{Hobbs {et~al.}(2005)Hobbs, Lorimer, Lyne, \& Kramer}]{Hobbs2005}
Hobbs, G., Lorimer, D.~R., Lyne, A.~G., \& Kramer, M. 2005, \mnras, 360, 974

\bibitem[{Hornschemeier {et~al.}(2005)Hornschemeier, Heckman, Ptak, Tremonti,
  \& Colbert}]{Hornschemeier2005}
Hornschemeier, A.~E., Heckman, T.~M., Ptak, A.~F., Tremonti, C.~A., \& Colbert,
  E.~J.~M. 2005, \aj, 129, 86

\bibitem[{Humphrey \& Buote(2008)}]{HB2008}
Humphrey, P.~J. \& Buote, D.~A. 2008, The Astrophysical Journal, 689, 983

\bibitem[{Hurley {et~al.}(2002)Hurley, Tout, \& Pols}]{Hurley2002}
Hurley, J.~R., Tout, C.~A., \& Pols, O.~R. 2002, \mnras, 329, 897

\bibitem[{Idiart {et~al.}(2007)Idiart, Silk, \&
  de~Freitas~Pacheco}]{Idiart2007}
Idiart, T.~P., Silk, J., \& de~Freitas~Pacheco, J.~A. 2007, \mnras, 381, 1711

\bibitem[{Irwin(2005)}]{Irwin2005}
Irwin, J.~A. 2005, The Astrophysical Journal, 631, 511

\bibitem[{Ivanova {et~al.}(2010)Ivanova, Chaichenets, Fregeau, Heinke,
  Lombardi, \& Woods}]{Ivanova2010}
Ivanova, N., Chaichenets, S., Fregeau, J., Heinke, C.~O., Lombardi, J. C.~J.,
  \& Woods, T.~E. 2010, The Astrophysical Journal, 717, 948

\bibitem[{Ivanova {et~al.}(2008)Ivanova, Heinke, Rasio, Belczynski, \&
  Fregeau}]{Ivanova2008}
Ivanova, N., Heinke, C.~O., Rasio, F.~A., Belczynski, K., \& Fregeau, J.~M.
  2008, \mnras, 386, 553

\bibitem[{Ivanova {et~al.}(2006)Ivanova, Heinke, Rasio, Taam, Belczynski, \&
  Fregeau}]{Ivanova2006}
Ivanova, N., Heinke, C.~O., Rasio, F.~A., Taam, R.~E., Belczynski, K., \&
  Fregeau, J. 2006, \mnras, 372, 1043

\bibitem[{Ivanova \& Taam(2003)}]{IT2003}
Ivanova, N. \& Taam, R.~E. 2003, \apj, 599, 516

\bibitem[{Justham \& Schawinski(2012)}]{JS2012}
Justham, S. \& Schawinski, K. 2012, \mnras, 423, 1641

\bibitem[{Kaaret {et~al.}(2011)Kaaret, Schmitt, \& Gorski}]{KSG2011}
Kaaret, P., Schmitt, J., \& Gorski, M. 2011, The Astrophysical Journal, 741, 10

\bibitem[{Kim \& Fabbiano(2010)}]{KF2010}
Kim, D.-W. \& Fabbiano, G. 2010, The Astrophysical Journal, 721, 1523

\bibitem[{Kobulnicky \& Fryer(2007)}]{KF2007}
Kobulnicky, H.~A. \& Fryer, C.~L. 2007, The Astrophysical Journal, 670, 747

\bibitem[{Kotani {et~al.}(2000)Kotani, Kawai, Nagase, Namiki, Sakano,
  Takeshima, Ueda, Yamaoka, \& Hjellming}]{Kotani2000}
Kotani, T., Kawai, N., Nagase, F., Namiki, M., Sakano, M., Takeshima, T., Ueda,
  Y., Yamaoka, K., \& Hjellming, R.~M. 2000, The Astrophysical Journal, 543,
  L133

\bibitem[{Kroupa(2001)}]{Kroupa2001}
Kroupa, P. 2001, \mnras, 322, 231

\bibitem[{Kroupa \& Weidner(2003)}]{KW2003}
Kroupa, P. \& Weidner, C. 2003, The Astrophysical Journal, 598, 1076

\bibitem[{Kubota {et~al.}(2005)Kubota, Ebisawa, Makishima, \&
  Nakazawa}]{Kubota2005}
Kubota, A., Ebisawa, K., Makishima, K., \& Nakazawa, K. 2005, The Astrophysical
  Journal, 631, 1062

\bibitem[{Lehmer {et~al.}(2010)Lehmer, Alexander, Bauer, Brandt, Goulding,
  {Jenkins, L. P.}, Ptak, \& Roberts}]{Lehmer2010}
Lehmer, B.~D., Alexander, D.~M., Bauer, F.~E., Brandt, W.~N., Goulding, A.~D.,
  {Jenkins, L. P.}, Ptak, A., \& Roberts, T.~P. 2010, The Astrophysical
  Journal, 724, 559

\bibitem[{Lehmer {et~al.}(2008)Lehmer, Brandt, Alexander, Bell, Hornschemeier,
  McIntosh, Bauer, Gilli, Mainieri, Schneider, Silverman, Steffen, Tozzi, \&
  Wolf}]{Lehmer2008}
Lehmer, B.~D., Brandt, W.~N., Alexander, D.~M., Bell, E.~F., Hornschemeier,
  A.~E., McIntosh, D.~H., Bauer, F.~E., Gilli, R., Mainieri, V., Schneider,
  D.~P., Silverman, J.~D., Steffen, A.~T., Tozzi, P., \& Wolf, C. 2008, The
  Astrophysical Journal, 681, 1163

\bibitem[{Lehmer {et~al.}(2007)Lehmer, Brandt, Alexander, Bell, McIntosh,
  Bauer, Hasinger, Mainieri, Miyaji, Schneider, \& Steffen}]{Lehmer2007}
Lehmer, B.~D., Brandt, W.~N., Alexander, D.~M., Bell, E.~F., McIntosh, D.~H.,
  Bauer, F.~E., Hasinger, G., Mainieri, V., Miyaji, T., Schneider, D.~P., \&
  Steffen, A.~T. 2007, The Astrophysical Journal, 657, 681

\bibitem[{Lehmer {et~al.}(2012)Lehmer, Xue, Brandt, Alexander, Bauer, Brusa,
  Comastri, Gilli, Hornschemeier, Luo, Paolillo, Ptak, Shemmer, Schneider,
  Tozzi, \& Vignali}]{Lehmer2012}
Lehmer, B.~D., Xue, Y.~Q., Brandt, W.~N., Alexander, D.~M., Bauer, F.~E.,
  Brusa, M., Comastri, A., Gilli, R., Hornschemeier, A.~E., Luo, B., Paolillo,
  M., Ptak, A., Shemmer, O., Schneider, D.~P., Tozzi, P., \& Vignali, C. 2012,
  The Astrophysical Journal, 752, 46

\bibitem[{Linden {et~al.}(2010)Linden, kalogera, Sepinsky, Prestwich, Zezas, \&
  Gallagher}]{Linden2010}
Linden, T., kalogera, V., Sepinsky, J.~F., Prestwich, A., Zezas, A., \&
  Gallagher, J.~S. 2010, The Astrophysical Journal, 725, 1984

\bibitem[{Linden {et~al.}(2009)Linden, Sepinsky, kalogera, \&
  Belczynski}]{Linden2009}
Linden, T., Sepinsky, J.~F., kalogera, V., \& Belczynski, K. 2009, The
  Astrophysical Journal, 699, 1573

\bibitem[{Luo {et~al.}(2012)Luo, Fabbiano, Fragos, Kim, Belczynski,
  Brassington, Pellegrini, Tzanavaris, Wang, \& Zezas}]{Luo2012}
Luo, B., Fabbiano, G., Fragos, T., Kim, D.-W., Belczynski, K., Brassington,
  N.~J., Pellegrini, S., Tzanavaris, P., Wang, J., \& Zezas, A. 2012, The
  Astrophysical Journal, 749, 130

\bibitem[{Marchesini {et~al.}(2009)Marchesini, van Dokkum,
  F{\"o}rster~Schreiber, Franx, Labb{\'e}, \& Wuyts}]{Marchesini2009}
Marchesini, D., van Dokkum, P.~G., F{\"o}rster~Schreiber, N.~M., Franx, M.,
  Labb{\'e}, I., \& Wuyts, S. 2009, The Astrophysical Journal, 701, 1765

\bibitem[{McClintock \& Remillard(2006)}]{McCR2006}
McClintock, J.~E. \& Remillard, R.~A. 2006, In: Compact stellar X-ray sources.
  Edited by Walter Lewin {\&} Michiel van der Klis. Cambridge Astrophysics
  Series, 157

\bibitem[{Mineo {et~al.}(2012{\natexlab{a}})Mineo, Gilfanov, \&
  Sunyaev}]{Mineo2012a}
Mineo, S., Gilfanov, M., \& Sunyaev, R. 2012{\natexlab{a}}, \mnras, 419, 2095

\bibitem[{Mineo {et~al.}(2012{\natexlab{b}})Mineo, Gilfanov, \&
  Sunyaev}]{Mineo2012b}
---. 2012{\natexlab{b}}, arXiv, 1207, 2157

\bibitem[{Mirabel {et~al.}(2011)Mirabel, Dijkstra, Laurent, Loeb, \&
  Pritchard}]{Mirabel2011}
Mirabel, I.~F., Dijkstra, M., Laurent, P., Loeb, A., \& Pritchard, J.~R. 2011,
  \aap, 528, A149+

\bibitem[{Pinsonneault \& Stanek(2006)}]{PS2006}
Pinsonneault, M.~H. \& Stanek, K.~Z. 2006, The Astrophysical Journal, 639, L67

\bibitem[{Podsiadlowski {et~al.}(2003)Podsiadlowski, Rappaport, \&
  Han}]{PRH2003}
Podsiadlowski, P., Rappaport, S., \& Han, Z. 2003, Monthly Notice of the Royal
  Astronomical Society, 341, 385

\bibitem[{Ptak {et~al.}(2007)Ptak, Mobasher, Hornschemeier, Bauer, \&
  Norman}]{Ptak2007}
Ptak, A., Mobasher, B., Hornschemeier, A., Bauer, F., \& Norman, C. 2007, The
  Astrophysical Journal, 667, 826

\bibitem[{Ranalli {et~al.}(2003)Ranalli, Comastri, \& Setti}]{Ranalli2003}
Ranalli, P., Comastri, A., \& Setti, G. 2003, \aap, 399, 39

\bibitem[{Reddy \& Steidel(2009)}]{RS2009}
Reddy, N.~A. \& Steidel, C.~C. 2009, The Astrophysical Journal, 692, 778

\bibitem[{Sana \& Evans(2011)}]{SE2011}
Sana, H. \& Evans, C.~J. 2011, Active OB stars: structure, 272, 474

\bibitem[{Scannapieco {et~al.}(2012)Scannapieco, Wadepuhl, Parry, Navarro,
  Jenkins, Springel, Teyssier, Carlson, Couchman, Crain, Dalla~Vecchia, Frenk,
  Kobayashi, Monaco, Murante, Okamoto, Quinn, Schaye, Stinson, Theuns, Wadsley,
  White, \& Woods}]{Scannapieco2012}
Scannapieco, C., Wadepuhl, M., Parry, O.~H., Navarro, J.~F., Jenkins, A.,
  Springel, V., Teyssier, R., Carlson, E., Couchman, H. M.~P., Crain, R.~A.,
  Dalla~Vecchia, C., Frenk, C.~S., Kobayashi, C., Monaco, P., Murante, G.,
  Okamoto, T., Quinn, T., Schaye, J., Stinson, G.~S., Theuns, T., Wadsley, J.,
  White, S.~D.~M., \& Woods, R. 2012, \mnras, 423, 1726

\bibitem[{Schiminovich {et~al.}(2005)Schiminovich, Ilbert, Arnouts, Milliard,
  Tresse, Le~F{\`e}vre, Treyer, Wyder, Budav{\'a}ri, Zucca, Zamorani, Martin,
  Adami, Arnaboldi, Bardelli, Barlow, Bianchi, Bolzonella, Bottini, Byun,
  Cappi, Contini, Charlot, Donas, Forster, Foucaud, Franzetti, Friedman,
  Garilli, Gavignaud, Guzzo, Heckman, Hoopes, Iovino, Jelinsky, Le~Brun, Lee,
  Maccagni, Madore, Malina, Marano, Marinoni, McCracken, Mazure, Meneux,
  Morrissey, Neff, Paltani, Pell{\`o}, Picat, Pollo, Pozzetti, Radovich, Rich,
  Scaramella, Scodeggio, Seibert, Siegmund, Small, Szalay, Vettolani, Welsh,
  Xu, \& Zanichelli}]{Schiminovich2005}
Schiminovich, D., Ilbert, O., Arnouts, S., Milliard, B., Tresse, L.,
  Le~F{\`e}vre, O., Treyer, M., Wyder, T.~K., Budav{\'a}ri, T., Zucca, E.,
  Zamorani, G., Martin, D.~C., Adami, C., Arnaboldi, M., Bardelli, S., Barlow,
  T., Bianchi, L., Bolzonella, M., Bottini, D., Byun, Y.-I., Cappi, A.,
  Contini, T., Charlot, S., Donas, J., Forster, K., Foucaud, S., Franzetti, P.,
  Friedman, P.~G., Garilli, B., Gavignaud, I., Guzzo, L., Heckman, T.~M.,
  Hoopes, C., Iovino, A., Jelinsky, P., Le~Brun, V., Lee, Y.-W., Maccagni, D.,
  Madore, B.~F., Malina, R., Marano, B., Marinoni, C., McCracken, H.~J.,
  Mazure, A., Meneux, B., Morrissey, P., Neff, S., Paltani, S., Pell{\`o}, R.,
  Picat, J.~P., Pollo, A., Pozzetti, L., Radovich, M., Rich, R.~M., Scaramella,
  R., Scodeggio, M., Seibert, M., Siegmund, O., Small, T., Szalay, A.~S.,
  Vettolani, G., Welsh, B., Xu, C.~K., \& Zanichelli, A. 2005, The
  Astrophysical Journal, 619, L47

\bibitem[{Symeonidis {et~al.}(2011)Symeonidis, Georgakakis, Seymour, Auld,
  Bock, Brisbin, Buat, Burgarella, Chanial, Clements, Cooray, Eales, Farrah,
  Franceschini, Glenn, Griffin, Hatziminaoglou, Ibar, Ivison, Mortier, Oliver,
  Page, Papageorgiou, Pearson, P{\'e}rez-Fournon, Pohlen, Rawlings, Raymond,
  Rodighiero, Roseboom, Rowan-Robinson, Scott, Smith, Tugwell, Vaccari, Vieira,
  Vigroux, Wang, \& Wright}]{Symeonidis2011}
Symeonidis, M., Georgakakis, A., Seymour, N., Auld, R., Bock, J., Brisbin, D.,
  Buat, V., Burgarella, D., Chanial, P., Clements, D.~L., Cooray, A., Eales,
  S., Farrah, D., Franceschini, A., Glenn, J., Griffin, M., Hatziminaoglou, E.,
  Ibar, E., Ivison, R.~J., Mortier, A. M.~J., Oliver, S.~J., Page, M.~J.,
  Papageorgiou, A., Pearson, C.~P., P{\'e}rez-Fournon, I., Pohlen, M.,
  Rawlings, J.~I., Raymond, G., Rodighiero, G., Roseboom, I.~G.,
  Rowan-Robinson, M., Scott, D., Smith, A.~J., Tugwell, K.~E., Vaccari, M.,
  Vieira, J.~D., Vigroux, L., Wang, L., \& Wright, G. 2011, \mnras, 417, 2239

\bibitem[{Tamura {et~al.}(2012)Tamura, Kubota, Yamada, Done, Kolehmainen, Ueda,
  \& Torii}]{Tamura2012}
Tamura, M., Kubota, A., Yamada, S., Done, C., Kolehmainen, M., Ueda, Y., \&
  Torii, S. 2012, The Astrophysical Journal, 753, 65

\bibitem[{Tremmel {et~al.}(2012)Tremmel, Fragos, Lehmer, Tzanavaris,
  Belczynski, kalogera, Basu-Zych, Farr, Hornschemeier, JENKINS, Ptak, \&
  Zezas}]{Tremmel2012}
Tremmel, M., Fragos, T., Lehmer, B.~D., Tzanavaris, P., Belczynski, K.,
  kalogera, V., Basu-Zych, A.~R., Farr, W.~M., Hornschemeier, A., JENKINS, L.,
  Ptak, A., \& Zezas, A. 2012, arXiv, 1210, 7185

\bibitem[{Tzanavaris \& Georgantopoulos(2008)}]{TG2008}
Tzanavaris, P. \& Georgantopoulos, I. 2008, \aap, 480, 663

\bibitem[{Valsecchi {et~al.}(2010)Valsecchi, Glebbeek, Farr, Fragos, Willems,
  Orosz, Liu, \& Kalogera}]{Valsecchi2010}
Valsecchi, F., Glebbeek, E., Farr, W.~M., Fragos, T., Willems, B., Orosz,
  J.~A., Liu, J., \& Kalogera, V. 2010, Nature, 468, 77

\bibitem[{Vattakunnel {et~al.}(2012)Vattakunnel, Tozzi, Matteucci, Padovani,
  Miller, Bonzini, Mainieri, Paolillo, Vincoletto, Brandt, Luo, Kellermann, \&
  Xue}]{Vattakunnel2012}
Vattakunnel, S., Tozzi, P., Matteucci, F., Padovani, P., Miller, N., Bonzini,
  M., Mainieri, V., Paolillo, M., Vincoletto, L., Brandt, W.~N., Luo, B.,
  Kellermann, K.~I., \& Xue, Y.~Q. 2012, \mnras, 420, 2190

\bibitem[{Voss \& Ajello(2010)}]{VA2010}
Voss, R. \& Ajello, M. 2010, The Astrophysical Journal, 721, 1843

\bibitem[{White \& Ghosh(1998)}]{WG1998}
White, N.~E. \& Ghosh, P. 1998, \apjl, 504, L31+

\bibitem[{Wu {et~al.}(2010)Wu, Yu, Li, Maccarone, \& Li}]{Wu2010}
Wu, Y.~X., Yu, W., Li, T.~P., Maccarone, T.~J., \& Li, X.~D. 2010, The
  Astrophysical Journal, 718, 620

\bibitem[{Xue {et~al.}(2011)Xue, Luo, Brandt, Bauer, Lehmer, Broos, Schneider,
  Alexander, Brusa, Comastri, Fabian, Gilli, Hasinger, Hornschemeier,
  Koekemoer, Liu, Mainieri, Paolillo, Rafferty, Rosati, Shemmer, Silverman,
  Smail, Tozzi, \& Vignali}]{Xue2011}
Xue, Y.~Q., Luo, B., Brandt, W.~N., Bauer, F.~E., Lehmer, B.~D., Broos, P.~S.,
  Schneider, D.~P., Alexander, D.~M., Brusa, M., Comastri, A., Fabian, A.~C.,
  Gilli, R., Hasinger, G., Hornschemeier, A.~E., Koekemoer, A., Liu, T.,
  Mainieri, V., Paolillo, M., Rafferty, D.~A., Rosati, P., Shemmer, O.,
  Silverman, J.~D., Smail, I., Tozzi, P., \& Vignali, C. 2011, \apjs, 195, 10

\bibitem[{Zhang {et~al.}(2012)Zhang, Gilfanov, \& Bogd{\'a}n}]{ZGA2012}
Zhang, Z., Gilfanov, M., \& Bogd{\'a}n, {\'A}. 2012, \aap, 546, 36

\bibitem[{Zuo \& Li(2011)}]{ZL2011}
Zuo, Z.-Y. \& Li, X.-D. 2011, The Astrophysical Journal, 733, 5

\end{thebibliography}
\end{document}